\documentclass[iop,apj,tighten]{emulateapj}
\usepackage{apjfonts} 
\usepackage{amsmath,amstext}
\usepackage[breaklinks,colorlinks,citecolor=blue,linkcolor=magenta]{hyperref} 
\usepackage[all]{hypcap} 
\usepackage{verbatim}

\begin{document}

\title{The Location and Environments of Neutron Star Mergers in an Evolving Universe}
\author{Brandon K. Wiggins\altaffilmark{1,2}, Christopher L. Fryer\altaffilmark{1}, Joseph M. Smidt\altaffilmark{1}, Dieter H. Hartmann\altaffilmark{3},\\ Nicole Lloyd-Ronning\altaffilmark{1}, Chris Belcynski\altaffilmark{4}}
\affil{$^{1}$Center for Theoretical Astrophysics, Los Alamos National Laboratory, Los Alamos, NM 87545}
\affil{$^{2}$Department of Physical Science,Southern Utah University, Cedar City, Utah 84721}
\affil{$^{3}$Department of Physics and Astronomy, Clemson University, Clemson, SC 29634-09781}

\affil{$^{4}$Nicolaus Copernicus Astronomical Center, Polish Academy of Sciences, ul. Bartycka 18, 00-716 Warsaw, Poland}

\altaffiltext{*}{wiggins@lanl.gov}

\begin{abstract}
The simultaneous detection of gravitational and electromagnetic waves from a binary neutron star merger has both solidified the link between neutron star mergers and short-duration gamma-ray bursts (GRBs) and demonstrated the ability of astronomers to follow-up the gravitational wave detection to place constraints on the ejecta from these mergers as well as the nature of the GRB engine and its surroundings.  As the sensitivity of aLIGO and VIRGO increases, it is likely that a growing number of such detections will occur in the next few years, leading to a sufficiently-large number of events to constrain the populations of these GRB events.  While long-duration GRBs originate from massive stars and thus are located near their stellar nurseries, binary neutron stars may merge on much longer timescales, and thus may have had time to migrate appreciably. The strength and character of the electromagnetic afterglow emission of binary neutron star mergers is a sensitive function of the circum-merger environment. Though the explosion sites of short GRBs have been explored in the literature, the question has yet to be fully addressed in its cosmological context. We present cosmological simulations following the evolution of a galaxy cluster including star formation combined with binary population synthesis models to self-consistently track the locations and environmental gas densities of compact binary merger sites throughout the cosmic web. We present probability distributions for densities as a function of redshift and discuss model sensitivity to population synthesis model assumptions. 
\end{abstract}

\maketitle

\section{Introduction}

The rapid variability of Gamma-Ray Bursts (GRBs) argues for engines associated with compact remnants.  Models for these bursts focused on scenarios invoking accretion onto, or evolution of, neutron stars (NS) or black holes (BH).  One particular class of models, black hole accretion disk engines, invokes rapid accretion via a wide range of scenarios ranging from stellar collapse (and the formation of these compact objects) and compact object mergers~\citep{woosley93,li98,popham99,fryer99}.  Observations of GRBs identify two primary subclasses based on the duration and spectral hardness of the prompt gamma-ray emission~\citep{kouveliotou93}.  Based on these observations, theoretical models argued that long GRBs are produced by the collapse of massive stars whereas short GRBs are produced by the merger of a binary consisting of either two NSs or a NS plus a BH~\citep{li98,fryer99,popham99,bulik99,Lee07,Osh08}.

The short life-times of massive stars in long GRB models argues that they must occur in star forming regions of actively star forming galaxies.  For NS-NS and NS-BH binaries, there is an additional delay as the orbital separation decays through gravitational wave emission.  This delay means that the merger, and its associated GRB, can occur long after star formation has ended in the GRB host galaxy.  Observations of the hosts of GRBs supported these concepts with long bursts typically occurring in star-forming galaxies whereas short bursts occurring in all galaxy types~\citep{gehrels05,berger05,fox05,davanzo09,tanvir13}.  

Neutron stars are believed to be born with some linear momentum (due to asymmetries in either the supernova ejecta mass or neutrinos~\citep[for a review, see][]{fryer06}), a.k.a. a "kick", during their formation and the effect of this kick has been studied for a range of compact object binaries~\cite[see, for example][]{Phi91,brandt95,fryer97,fryer98,Por98,Bethe98}.  Because kicks are imparted on these compact remnants at birth, merger models predict that short GRBs should occur offset from the star forming birth site, and in fact even from the host galaxy.  Theory predicted offset distributions for these mergers~\citep{bloom99,fryer99,belczynski02,voss03,belczynski06,kelley10,behroozi14} that were ultimately confirmed by observations~\citep{troja08,fong13}.  This confirmation firmly established these merger models as the leading paradigm for short GRBs, albeit on the basis of a statistucal argument.  The detection of a GRB associated with an aLIGO~\citep{Ab17} neutron star merger event further supports the theory that short-duration GRBs (sGRBs) are produced by the merger of two neutron stars, and does so in a more direct way.  This detection also shows the potential role electromagnetic counterparts can play in probing NS binary circum-merger environments.   

For long-duration GRBs produced in the collapse of massive stars, the surrounding density (determined both by the star formation region and strong stellar winds) is believed to be high.  But the offsets of short GRBs suggests a wide range of densities, trending toward lower values relative to long-GRB environments.  Observations confirmed this range, placing constraints on the density of the circumstellar medium around short GRBs~\citep{fong13}. Because GRB redshifts are not small in general, cosmological evolution of the gas associated with galaxies and the circum-galactic medium should be taken into account when addressing questions about short-GRB environments. 

Coupling cosmology calculations with neutron star merger results has recently allowed scientists to conduct increasingly detailed studies of the sites of GRB events~\citep{zemp09,kelley10,behroozi14}.
In this paper, we present simulations following the evolution of a galactic cluster from high redshift to the present-day universe with a prescription for star formation. We post-process these calculations with  binary population synthesis models that provide the fraction of massive stars that form merger neutron star binaries along with the characteristics of these binaries which allow us to estimate the merger rate as a function of redshift and the distribution of merger offsets and circum-merger densities of these bursts.  These results allow us to calculate the radio signals from neutron star mergers. We discuss the cosmological models and the implementation of the population synthesis results in section~\ref{sec:simulations}.  The properties of neutron star mergers are discussed in section~\ref{sec:results} and the resultant radio signals are discussed in section~\ref{sec:radio}.  We conclude with a review and discussion of our calculation uncertainties. 

\section{Simulations}
\label{sec:simulations}

Our calculations combine cosmological simulations with a post-process using the results of population synthesis models to predict the locations and circum-merger densities of NS binaries. We describe the cosmological simulations in Section~\ref{sec:cosmological}, the population synthesis calculations used in this study in Section~\ref{sec:popsynth} and our integration of the two in Section~\ref{sec:integ}.

\subsection{Cosmological Simulations}
\label{sec:cosmological}

\begin{figure*}
\label{fig:simulation}
\begin{center}
\includegraphics[width=\textwidth]{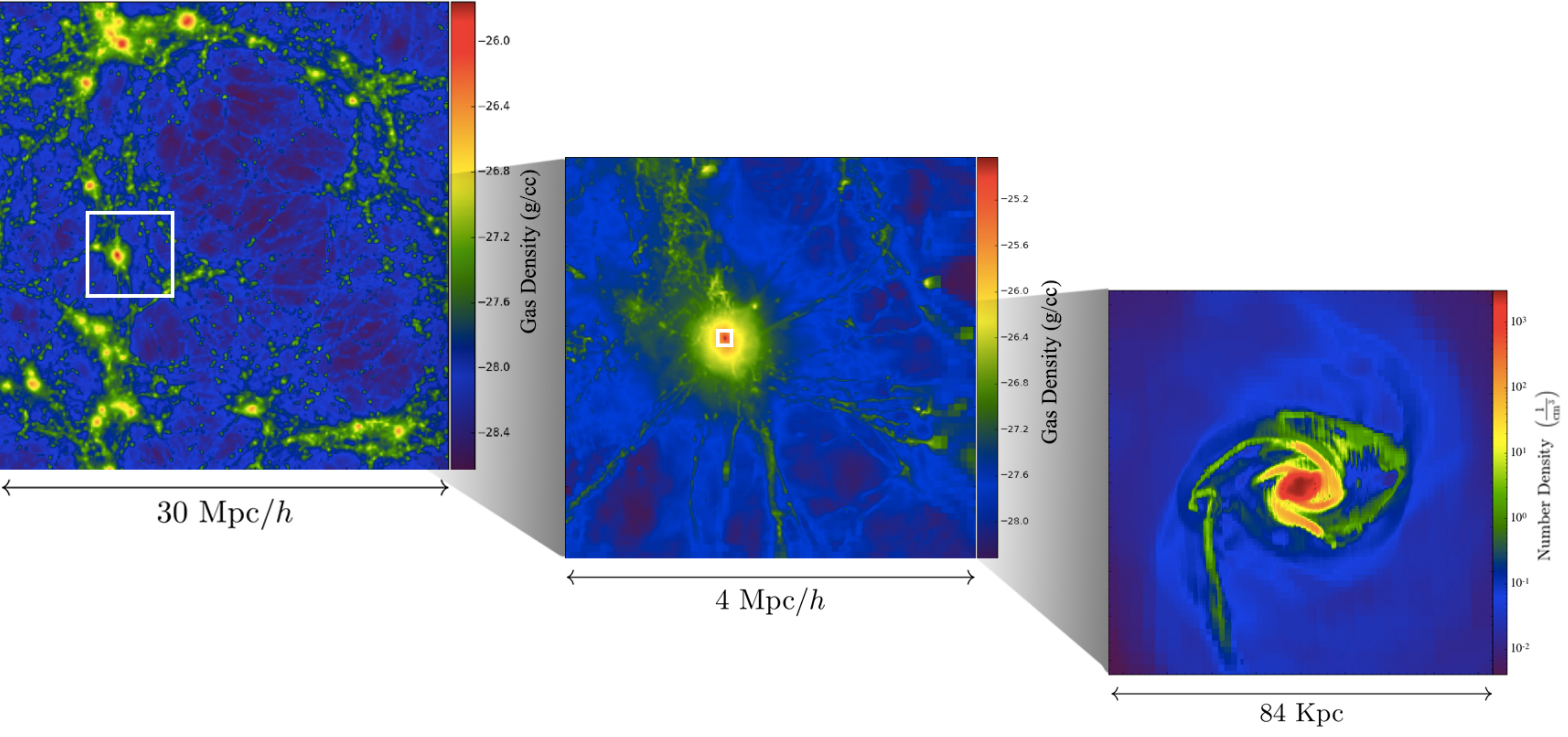}
\end{center}
\caption{Projection plots of gas density of our cosmological calculations at various scales. Our initial cosmological calculations spanned a $\sim 30$ Mpc co-moving volume. A subsequent deep dive calculation of a (4 Mpc/$h$)$^3$ box with eleven levels of refinement and primordial chemistry on the largest halo was then carried out. On refinement deep dives, our resolution is sufficient to resolve galaxy structure in this cluster.}
\end{figure*}


In this paper, the cosmological simulations were preformed using {\it Enzo} \citep{bryan14}, a cosmology code that contains the necessary modules for hydrodynamics, radiation, primordial chemistry and black hole physics required in this calculation. The initial conditions were generated by MUSIC \citep{hahn11} using the Planck 2015 TT,TE,EE+lowP+lensing+ext best fit parameters (Planck collaboration).  MUSIC was run using the \cite{eisenstein98} transfer function with second-order perturbation theory enabled and an initial redshift of $z = 200$. MUSIC seeded both the initial dark matter and baryon position, density and velocity distributions to be run in $Enzo$ with periodic boundary conditions. The initial box-size was 30 Mpc/h with 256$^3$ spatial resolution on the root grid.

These initial dark matter plus baryon initial conditions were evolved in {\it Enzo} down to redshift $z = 0$ without refinement. At that point, a halo finder using the HOP algorthim~\citep{skory10} was used to find a dark matter halo with mass sufficient to form a cluster of galaxies.  The simulation was then rerun with eleven levels of refinement across a 4 Mpc/h window enclosing the desired halo with 9-species primordial chemistry and cooling (H, H+, He, He+, He++, e-, H2, H2+ and H-) to get the correct baryon physics at small scales. The refinement for the fine-grid region was triggered by overdensities in dark matter and baryons with the additional criteria of requiring 32 cells across a Jean's length. Eleven levels of refinement represents a smallest spacial resolution of $\sim $ 57 pc/$h$, a couple orders of magnitude smaller than the baryon disk size. An image of this disk is shown in Figure ~\ref{fig:simulation}. 

The final volume we refine on ($\sim$4 Mpc/h) for our high-resolution cosmological simulation is not large enough to create a sufficient number of galaxies in varieties, sizes and with relative spatial distributions to be representative of galaxies in the entire universe. Our choice in the highly-resolved volume is motivated by practical memory considerations with the requirement that our simulation resolve scales with number densities $n > 10^3$ cm$^{-3}$ in the galaxies formed. \textit{Enzo}'s AMR data structure is memory intensive, rendering refinement of multiple halos to the desired number densities a practical impossibility.

This high-resolution simulation produced a series of time dependent data-dumps which follow the evolution of the halo and formation of the galaxy cluster as a function of redshift. These data were used in the post-process to determine the location of DNS merger locations (see \S 2.3). For visualization in this paper, we used YT (Turk et al. 2011) with techniques described in Smidt et al. 2016. 

Star formation is included using the \texttt{StarParticleCreation = 2} option in {\it Enzo} which immediately deposits supernova thermal energy and metals into the zone where stars were initially created instead of following the trajectory of star/star cluster particles. With this option, {\it Enzo} can calculate star formation rates using a global Schmidt Law \citep{K2003} or a method by \cite{cen92}. Our results in this paper are possibly sensitive to the star formation prescription and so we compare our results with those of another prescription in section 5. By simplifying our treatment of star formation, we achieve higher levels of refinement to better resolve the gas density of scales were DNSs are likely to merge.

\subsection{Population Synthesis Models}
\label{sec:popsynth}

Although a large fraction of massive stars form binaries, only a small fraction of these binaries remain bound through two supernova explosions to form neutron star binaries.  The rate of formation as well as the distributions of velocities and merger times for our calculations use the results from two different population synthesis calculations:  the StarTrack population synthesis code~(\cite{belczynski02b,belczynski08}) and the Fryer population synthesis code~(\cite{fryer98},\cite{fryer99}).  Both codes produce roughly the same range of rates of neutron star mergers in the Milky Way:  0.01-100\,Myr$^{-1}$ with most simulations producing rates of 1-10\,Myr$^{-1}$.  This agrees with many of the population synthesis studies\citep[e.g.][]{Por98,voss03}, but the high end of these rates are needed to match the current inferred rates from the detection of GW170817~\citep{belczynski17}.  Uncertainties in stellar evolution and supernovae can alter the characteristics of the neutron star binaries produced and these uncertainties are discussed in a number of population studies, e.g. \citep{fryer98,fryer99,dominik12}.  Our choice of simulations from both the StarTrack synthetic universe calculations and the Fryer models produce a broad range of possible binary conditions due to these uncertainties.

The StarTrack code is based on revised formulas from \cite{hurley00}; updated with new wind mass loss prescriptions, calibrated tidal interactions, physical estimation of the donor's binding energy ($\lambda$) in common envelope calculations and convection driven, neutrino enhanced supernova engines. A full description of these updates is given in \cite{dominik12}.  In the standard model, \cite{dominik12} use the derived binding energies from \cite{xu10} and a maximum neutron star mass of $2.5\,M_\odot$.  The compact remnant mass distribution is determined using the prescription from\citep{fryer2012}.  Neutron star kick distributions are typically inferred from pulsar velocity distributions, and a number of analyses exist~\citep{cordes98,arzoumanian02,hobbs05}.  These kick distributions include both single and two-component Maxwellian fits to the data.  In our StarTrack simulations, we employ a single Maxwellian ($\sigma=265$ km s$^{-1}$).  We use the models in the synthetic universe data, with our model numbers corresponding to the model variants in this database (https://www.syntheticuniverse.org). We test sensitivity to uncertainty in the population synthesis on NS binary merger location and density by carrying out our postprocess multiple times with 7 StarTrak variations which were representative of the models in Dominik et al. (2012). These models are summarized in Table\ref{tab:startrack}.

\capstartfalse
\begin{deluxetable*}{cclccc} 
\tabletypesize{\footnotesize} 
\tablecolumns{4} 
\tablewidth{0pt} 
\tablecaption{ StarTrack Model Descriptions \label{table:results1}} 
\tablehead{ 
\colhead{Model} & \colhead{Dominik et al. (2012) Model Name} & \colhead{Parameter Varied} & \multicolumn{3}{c}{Delay Time Range} \\ & & & \colhead{$Z_\odot$} & &\colhead{$10^{-1} Z_\odot$} } 
\startdata 
StarTrack 1 & Var 1 & $\lambda = 0.01$, common envelope very tightly bound & 52 Myr &&  82 Myr\\
StarTrack 2 & Var 2 & $\lambda = 0.1$, common envelope tightly bound &  $90-618$  Myr && $0.2-1.6$ Gyr\\
StarTrack 3 & Var 3 & $\lambda = 1.0$, common envelope less tightly bound &  $1.2-2.2$ Gyr&& $0.9-2.4$ Gyr\\
StarTrack 4 & Var 5 & varied max NS mass &  $1.2-2.2$ Gyr &&  $0.9-2.4$ Gyr\\
StarTrack 5 & Var 7 & $\sigma = 132.5$ km/s &  $1.3-2.2$ Gyr &&  $1.0-2.17$ Gyr\\
StarTrack 6 & Var 11 & divide mass transfer by 2 &  $1.0-1.6$ Gyr &&  $1.0-2.3$ Gyr\\
StarTrack 7 & Var 15 & allow $\lambda$ to vary & $0.8-1.6$ Gyr &&  $0.4-2.1$ Gyr
\enddata 
\label{tab:startrack}
\end{deluxetable*}
\capstarttrue

One of the uncertainties not studied extensively in the the synthetic universe calculations is the stellar radius for giant stars.  This was addressed by the calculations using the Fryer population synthesis code that uses fits of stellar radii~(\cite{kalogera98}) with a parameterization of the formula (allowing larger or smaller stellar radii) to determine the rate-dependence on the radius prescription (in short, fewer close binaries are made if the radii are smaller, but the binaries formed are tight binaries).  Another difference between this code and StarTrack is that this code uses a 2-component Maxwellian kick distribution (40\% of stars with a $\sigma=90 \, {\rm km s^{-1}}$ and 60\% of stars with a $\sigma=350 \, {\rm km s^{-1}}$ based on fits to both pulsar studies~\citep{arzoumanian02} and comparisons to globular cluster populations~\citep{fryer97}.  The most important difference is that, for this paper, the results from the Fryer simulations focus on models that do not allow the survival of neutron star systems that undergo a NS common envelope phase.  This extreme case would occur only if hypercritical accretion causes the neutron star to collapse to a black hole, considered unlikely in most common envelope scenarios~\citep{macleod15}.  We use a range of population synthesis from StarTrack~\citep{dominik12} and Fryer codes to span the space of possible population synthesis parameters.  

\begin{figure}[t]
	\includegraphics[width=\columnwidth]{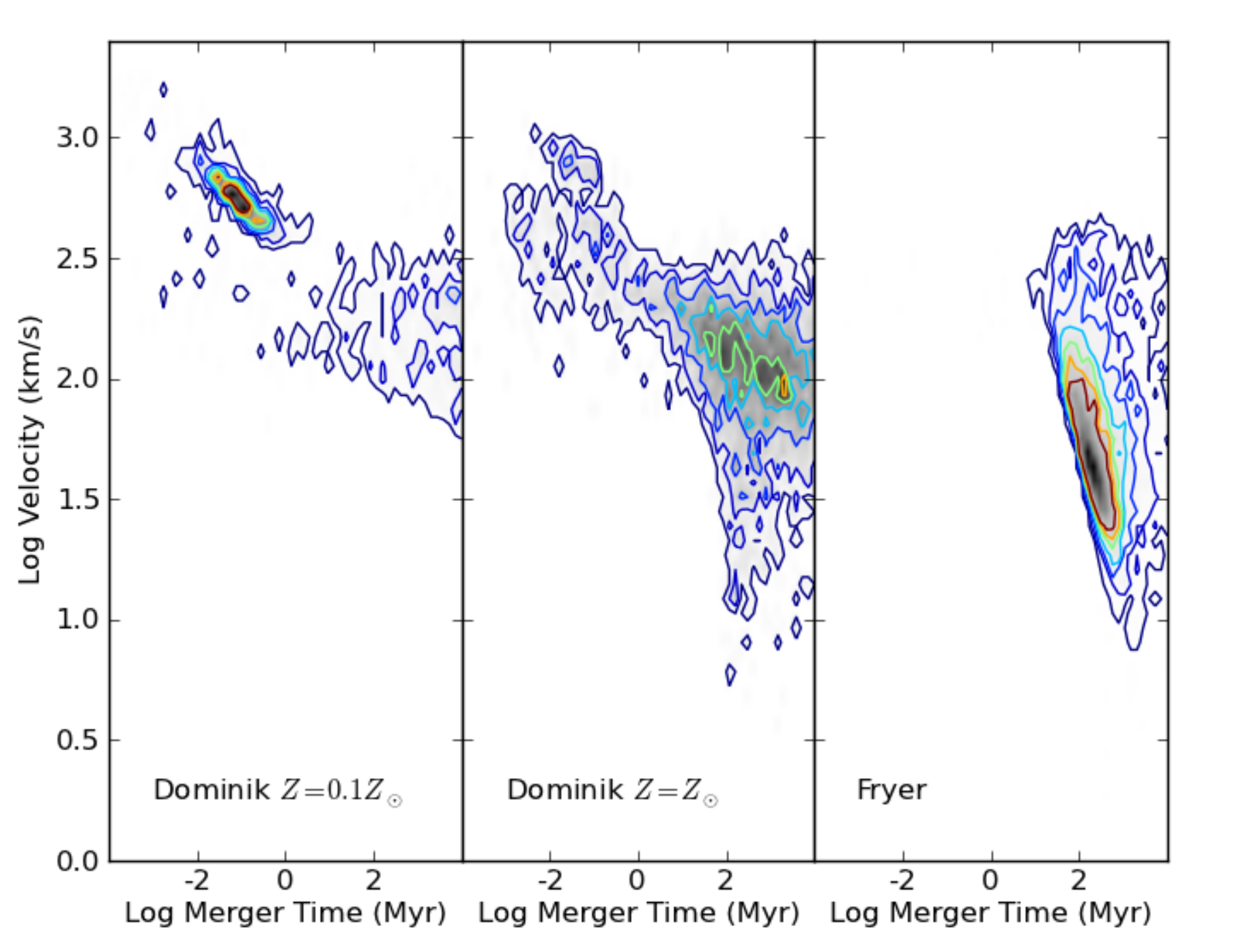}
    \caption{Merger velocities versus merger times after the formation of the second neutron star for individual systems produced in 3 different population synthesis models.  The "Dominik" models are the results from the Standard model in~\citep{dominik12} using the StarTrak code.  These two results refer to the solar and 1/10th solar variants of this standard model.  The Fryer model is produced by the population synthesis code described in \citep{fryer99}.  The number of short-merger time systems ($<1$\,Myr) can vary dramatically with population synthesis parameters and code assumptions.  We neglect the motion between the formation of the first and second neutron star, but we see from figure~\ref{fig:vsrs} that this aspect of the motion will not dramatically affect our results.}
    \label{fig:vmtm}
\end{figure}

The systemic velocity and merger times as measured from the time of the formation of the second neutron star depend on both the initial conditions (e.g. metallicity) and uncertain parameters in population synthesis calculation.  Figure~\ref{fig:vmtm} shows the distribution of merger velocities versus merger times for two metallicities (solar, 1/10th solar) for the standard model from~\citep{dominik12} as well as a sample simulation using a bimodal kick distribution from the Fryer code~\citep{fryer99}.  The bulk of the systems have merger times greater than 10\,Myr but some systems have merger times less than 1\,Myr. The fraction of these short merger systems range from $<$1\% to up to 15-20\%.  The distribution of velocities, the number of low-velocity systems, the distribution of merger times and the shortest merger can vary dramatically depending on the characteristics of the simulations.  The expansion of a star in the giant phase is less dramatic for lower metallicity systems.  For these lower-metallicity systems to undergo mass transfer or a common envelope phase (required to make the extremely tight orbits to produced binaries that will merge in a Hubble time), they must be in tighter orbits.  These tighter orbits, in turn, produce tighter binaries, a large fraction of which will undergo a helium giant common envelop phase producing extremely tight orbits with fast merger times and large systematic velocities.  Hence, in the low-metallicity systems, we produce a larger fraction of short merger-period systems.

Although there is a basic trend that higher metallicity simulations produce more low-velocity systems, the number of these systems also depend upon the kicks and the binding energy. Because halo metallicity increases with diminishing redshift and NS binaries formed from higher metallicity environments have on average, larger merging times, one might assume that aLIGO/VIRGO would more frequently detect NS binaries born from higher-metallicity environments. In actuality, this depends on the star formation and metal enrichment history of the host galaxy: if the galaxies are well-enriched with metals before peak of star formation ($\sim z =2$), we would indeed expect NS binaries from more metal-enriched SRFs (e.g. Figure 2, middle panel) to dominate detections.

\begin{figure}[t]
	\includegraphics[width=\columnwidth]{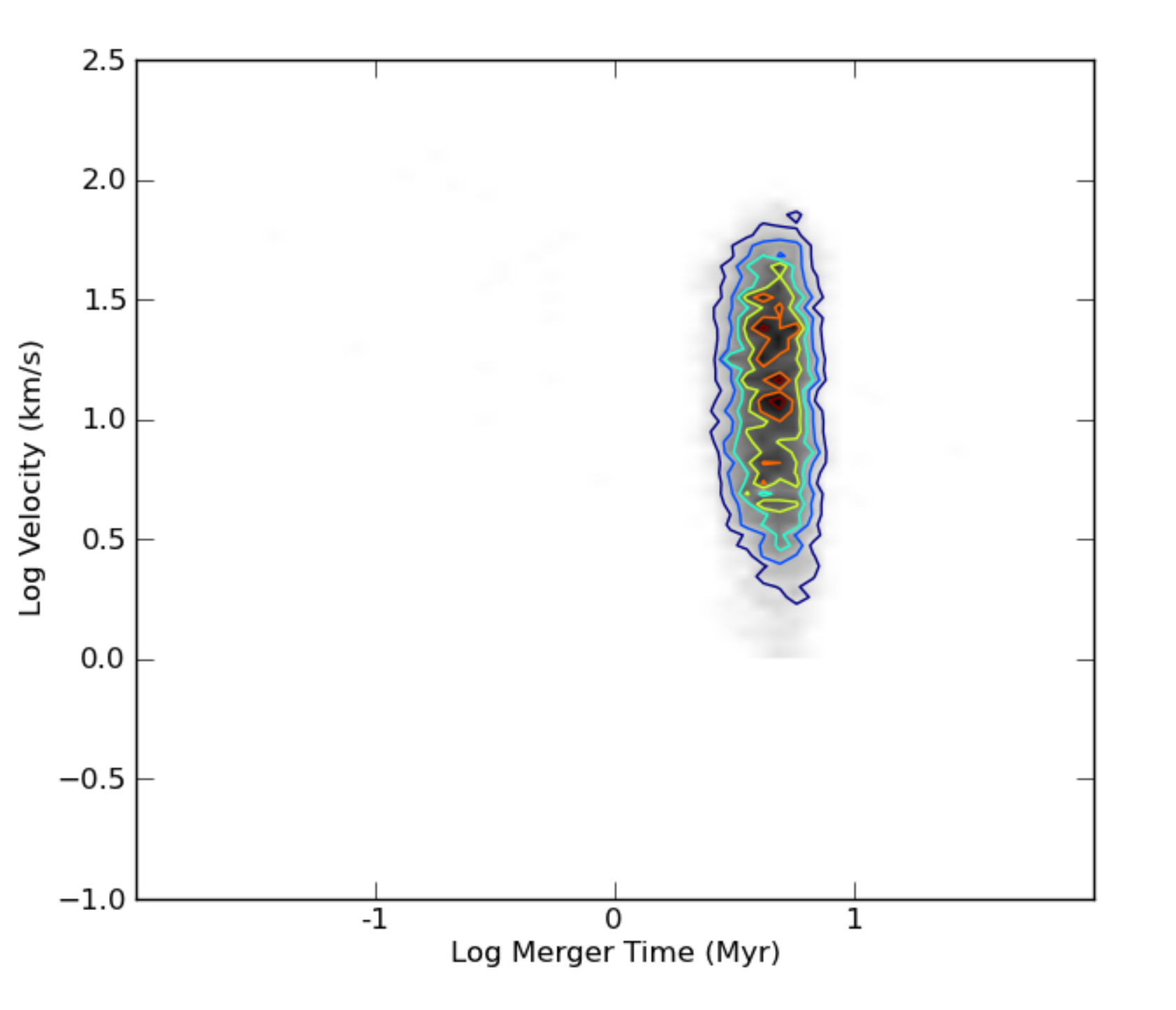}
    \caption{Velocity of the system after the formation of the first neutron star and before the formation of the second neutron star system versus the time between these two events for the Fryer model shown in figure~\ref{fig:vmtm}.  The position of the system after the merger should include both phases but, comparing these velocities and times to their comparable model in figure~\ref{fig:vmtm}, we see the affect of this phase is minimal.}
    \label{fig:vsrs}
\end{figure}

To truly determine the motion of the binary systems, we must also include both the motion in the period between the formation of the first neutron star and the second neutron star.  The velocities of these systems are typically lower (figure~\ref{fig:vsrs}) and typical times between the formation of the first neutron star and the second is between 1-10\,Myr.  The low velocity and modest evolution time for this phase in the evolution means that this phase unlikely to play a large role in the final location of the merging system.

\subsection{Integration of NS binary orbits}
\label{sec:integ}

Enzo contains a star class which can be used to integrate orbits of stars created. Unfortunately, Enzo distributes a copy of all star particles to all processors, which can lead to memory issues in calculations with many levels of refinement or large numbers of stars. In addition, Enzo distinguishes the creation of Pop III stars from Pop II and Pop I stellar clusters, a single option being available for the duration of a calculation. As mentioned previously, in this paper, we used \texttt{StarParticleCreation = 2} which only deposits thermal energy and metals locally based on star formation rate (SFR) and leave remaining aspects of NS binary formation to our post-process. Because the SFR as a function of position is not reported within data dumps, we infer the SFR during post-process with the method from \cite{cen92}. 

NS binary production is a strong function of local metallicity and aLIGO will detect compact object mergers originating from high and low metallicity environments. In principle, we should adopt NS binary production rates and characteristics based on local gas metallicity. For the scope of our work, we adopt a crude, two-metallicity production rate~\citep[e.g.][]{DMB2015} based on the metallicities of our two Star Track metallicities ($Z_\odot, \hspace{3 pt} 10^{-1} Z_\odot$). This simplified approach still allows us to study the sensitivity of our results to metallicity. A thorough treatment of NS production rate would require a fine grid of population synthesis models which were not available for the models employed in this paper. Higher metallicity population synthesis calculations robustly predict more NS-NS binaries. Between the two Star Track models we study in this paper, NS-NS binary formation is larger in higher-metallicity models by a factor of $\sim 3-4$ for a fixed SFR. We adopt the NS binary production rate $\mathcal{R}_{\mbox{\tiny DNS}}$ for a model with metallicity closest that of the gas, i.e.

\begin{equation}
\mathcal{R}_{\mbox{\tiny DNS}} =
\begin{cases}
0.001 \mbox{ SFR } & \mbox{ if } Z/Z_\odot > 10^{-0.5}\\
0.00025 \mbox{ SFR } & \mbox{ if } Z/Z_\odot < 10^{-0.5}.
\end{cases}
\end{equation}
where we have chosen 0.1\% of SFR for NS binary production for solar metallicity (e.g. Chruslinska et al. 2017a). This simplification does not treat Population III stars whose smaller SFR per comoving volume and top-heavy IMF and, hence, smaller fraction of ZAMS binaries~\citep{belczynski17b} would render their odds of detection smaller than other populations.

Binaries containing massive stars which are the progenitors of NS binaries undergo a series of natal kicks arising from supernovae of their members. Because of the relatively short stellar lifetimes (a few Myr), the NS binary is assumed to be created at the time of star formation. Each binary system is modeled as a single particle with initial velocities equal to the local gas velocity added to the natal kick velocity obtained from population synthesis models (see section 3.2). The dynamics of these binaries are given by Newton's equations for comoving coordinates i.e.

\begin{eqnarray}
\frac{d\textbf{x}}{dt} &=& \frac{1}{a} \textbf{v}\\
\frac{d \dot{\textbf{x}}}{dt} &=& -\frac{\dot{a}}{a} \dot{\textbf{x}} - \frac{1}{a}\nabla \phi
\end{eqnarray}
for $\textbf{x}$ and $\dot{\textbf{x}}$ are position and velocity with $a$ and $\phi$ being the scale factor and gravitational potential respectively.

Because our NS binaries merge on timescales $\sim 10^3$\,Myr, galactic and cosmological potentials may evolve substantially during the lifetime of the NS binary. The cosmological potential is sampled at discrete moments in time, the gravitational potential is linearly interpolated between dumps throughout the orbit integration. The orbit is integrated up until the moment of merging, with a merger timescale taken again from population synthesis calculations. 

We integrate NS binary orbits with the Verlet integrator with fixed timestep sizes. We carried out a serious of simulations with successive reductions in global timestep size until distributions in this paper converged. To check the accuracy of our scheme, we also integrated a set of NS binaries with a fourth order Runge-Kutta (RK4) scheme and timestep control and found our distributions to be robust. At the moment of merging of an NS pair, the local density, offset from origin galaxy, and redshift of the merger are recorded. 

It is important to emphasize that natal kick velocities and merger times for each NS binary are not drawn from independent distributions. Each star in our calculation represents a distinct NS binary system generated by population synthesis models, each with their own kick and merger timescale. In Figure 2 we plot the dependency of merger time on natal kick velocity to illustrate the correlation between quantities for a given NS binary. 

The effects of energy drift are not trivial to explicitly track for our NS binaries because their background potential evolves. Energy drift, a symptom of a timestep size too large for a particular position of a body on an orbit, would bias the predicted merger locations of the NS binaries. Energy drift in our calculations could potentially arise from very steep potential gradients in very dense regions on the simulation domain. To test for the presence of energy drift, we evolved NS binary positions in a static cosmological potential for $\sim 10$ Gyr. In such tests, we find that we conserve energy to 1\%. Maximum speeds attained by NS binaries in our calculations  $\gtrsim 1000$ km/s consistent with very high-velocity pulsars.

\section{Results}
\label{sec:results}

\subsection{Location}
\label{sec:location}

\begin{figure}
\includegraphics[width=\columnwidth]{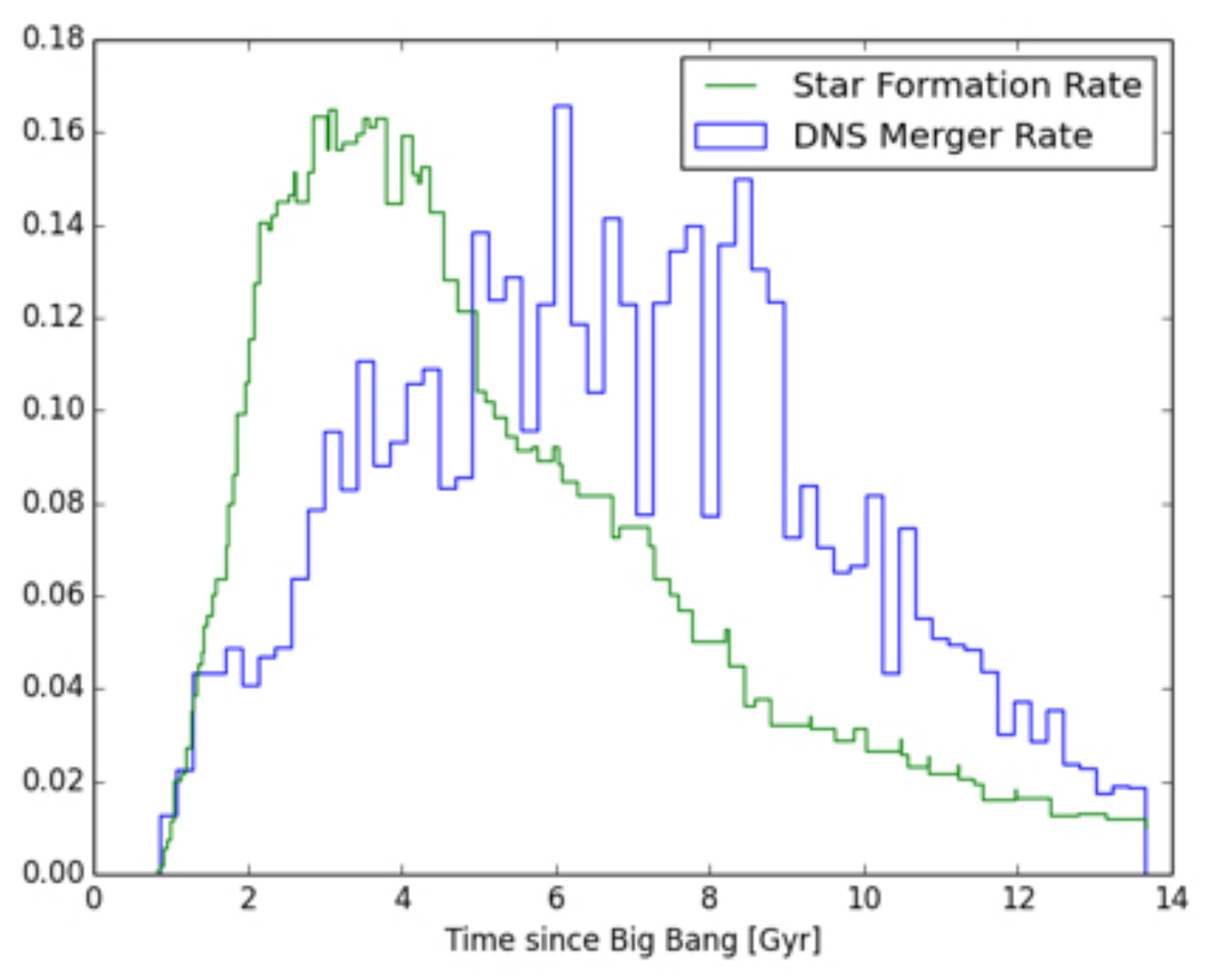}
\caption{Normalized star formation rate from ENZO calculations against normalized NS binary merger rate. Because of substantial delays in binary NS mergers, the peaks are offset by several Gyr.}
\label{sfr}
\end{figure}

The relative rates of star formation in our simulations correspond well to observations (see Figure \ref{sfr}) with star formation rate (SFR) peaking around 3 Gyr ($z=2$) post-Big Bang. The predicted peak of NS binary mergers is delayed from peak of star formation as expected, peaking around 6 to 8 Gyr following the Big Bang ($z = 0.6 -1.0$) which is similar to observations of sGRBs for which redshift data exist (Fong et al. 2015). A K-S test between observed sGRB times (Fong et al. 2015) and our theoretical temporal NS binary merger distribution is unable to reject the hypothesis that the observed redshifts are drawn from our theoretical distribution at 95\% confidence. While this test does not take into account left-censoring of sGRB data and neglects sGRBs without measured redshift, the correlation demonstrates the calculations' agreement with observations.

Our high-resolution models resolve galactic scales including spiral disk structure (see Figure \ref{fig:simulation}). In these demanding calculations, we do not self-consistently model the detailed hydrodynamics of supernova remnants and mixing of metals in the interstellar medium, but rather deposit metals and thermal energy locally based on star formation rate. Even with refinement this practice can lead to an underestimate of the amount of gas driven from a halo from supernovae because the thermal energy content of a supernova may be deposited over a cell with a volume much larger than a supernova remnant. Modeling supernova remnants explicitly is the best practice to constrain how much gas has been driven from a halo but was computationally infeasible with the already ambitious character of our calculation. Supernova blasts can act to drive gas from halos at very early times when potential wells are more shallow and lead to more chaotic galaxy structure for smaller dwarf galaxies. We calculate physical projected offsets $\delta$R in kpc for NS binaries originating from roughly Milky Way-sized ($\sim 10^{12}$\,M$_\odot$ in baryons + dark matter) galaxies in such a run, post-processed with NS orbits informed by natal kick velocities and delays from our population synthesis models.  

\begin{figure*}
\begin{center}
\includegraphics[width=340 pt]{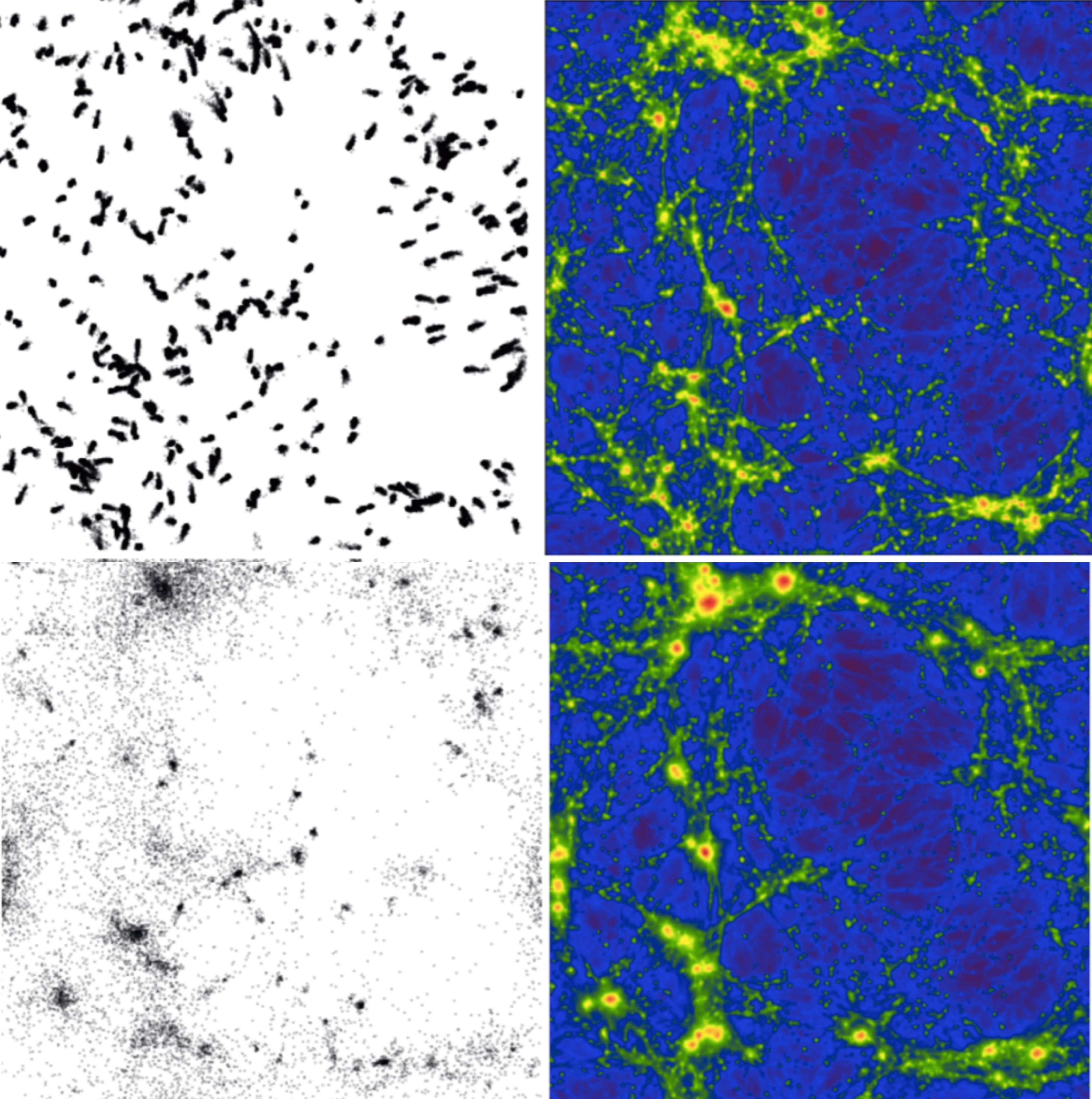}
\end{center}
\caption{\textit{Left Panels:} Projection of NS Binary merger locations on very large scales, i.e. a co-moving $\sim$30 Mpc box integrated from $z=2$ to z = 3.5 ($\sim$ 10 - 12 Gyr lookback time, top-left panel) and from $z = 0$ to $z = 0.2$ ($\sim 0 - 3$ Gyr lookback time, bottom-left panel). \textit{Right Panels:} Projection of gas density. Note that near peak of star formation ($z = 2$) a large population of NS binaries merge near their SFRs. At low redshift, NS binaries produced during this early spurt of star formation with larger merger timescales have had time to wander from their hosts.}
\label{fig:cos1}
\end{figure*}

Figure~\ref{fig:cos1} shows the locations of the neutron star binaries when they merge for a 40\,Mpc box (co-moving) at a redshift of 0 and 2.  For the most part, these mergers occur near the gas distribution at these large scales, but the locations of the neutron star mergers extend well beyond the highest concentration of gas.  Although there is a paucity of mergers occurring in the inter-cluster voids, the mergers are far more widespread than the gas in these cosmological models.  

\begin{figure}
\includegraphics[width=\columnwidth]{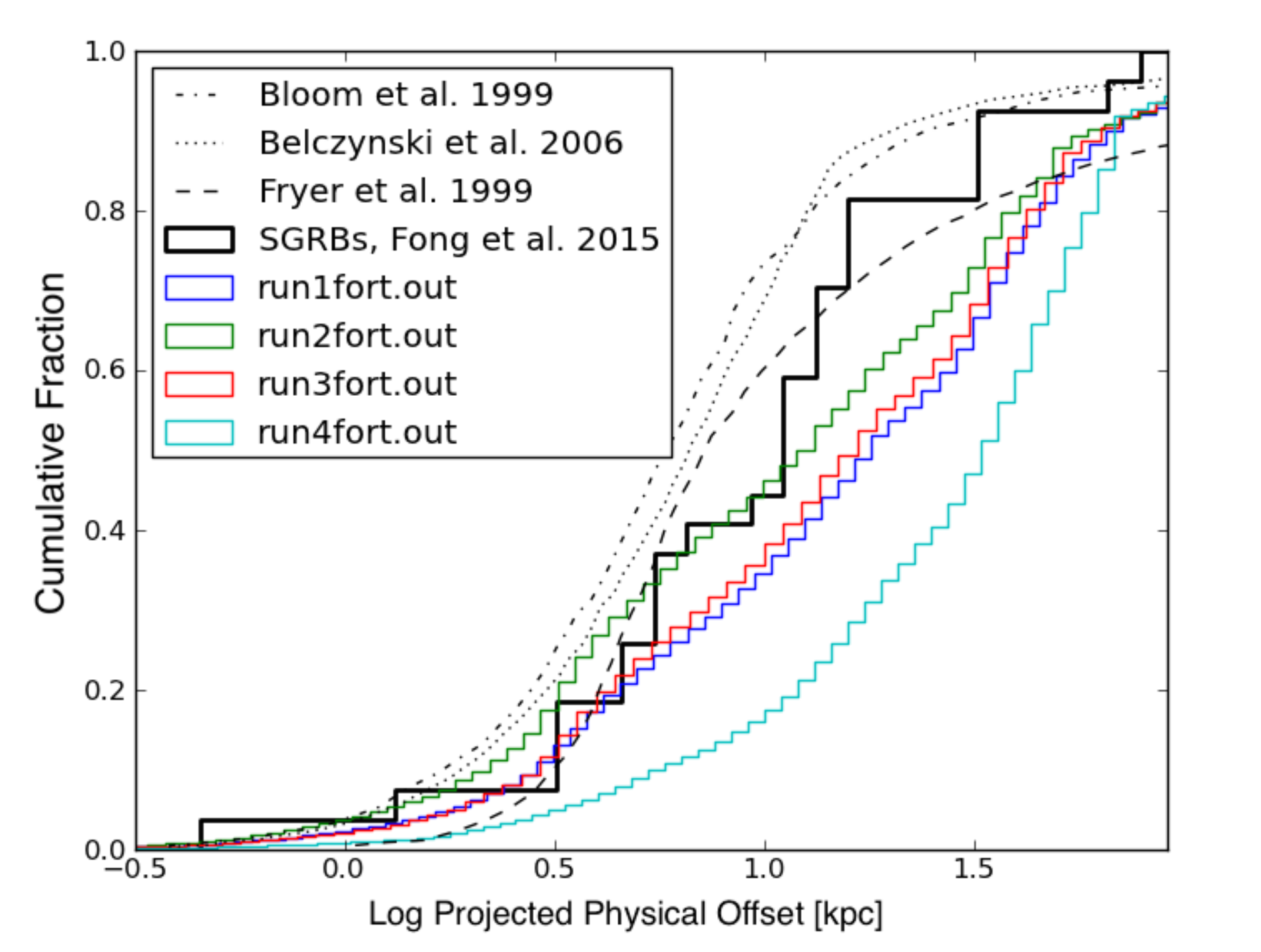}
\includegraphics[width=\columnwidth]{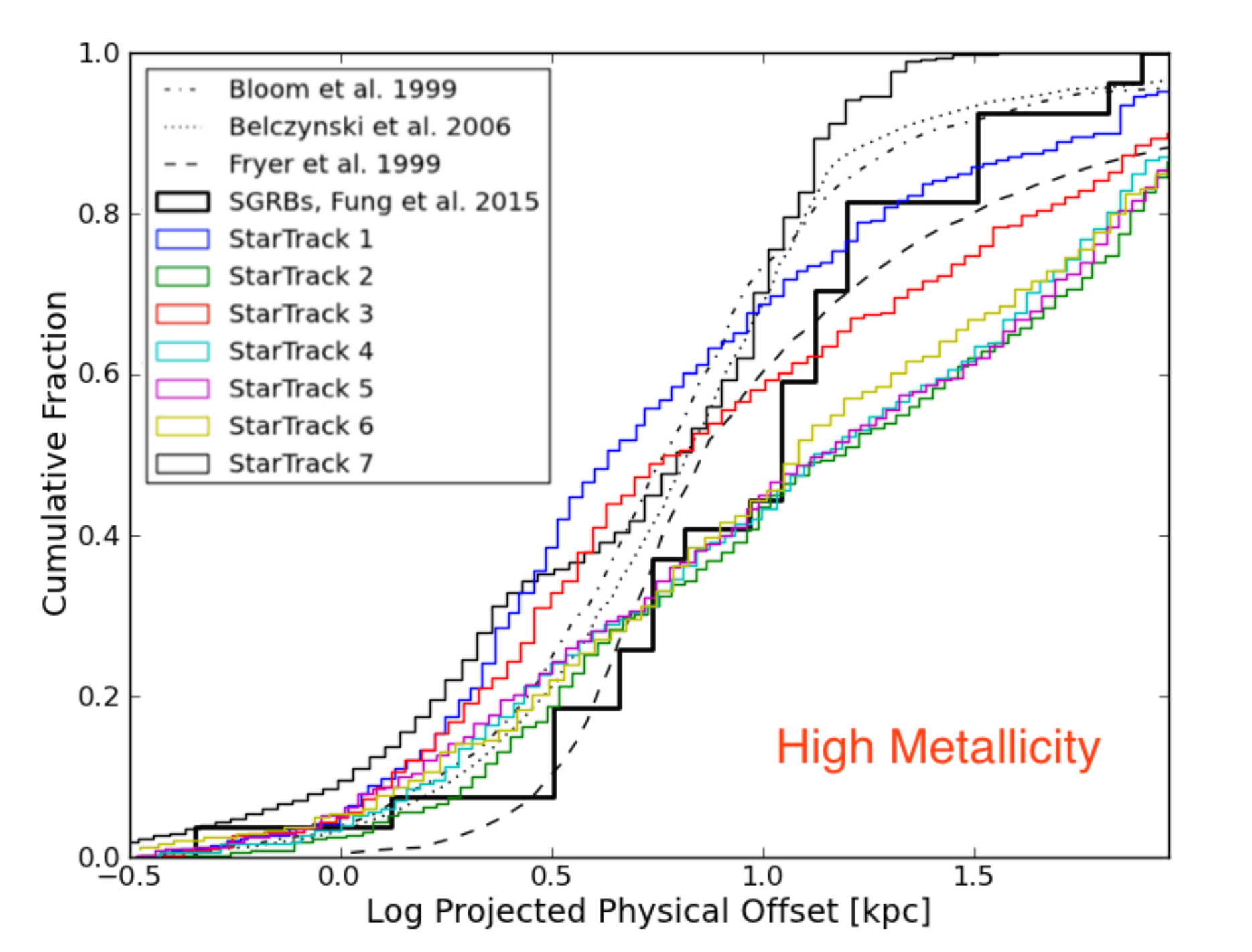}
\includegraphics[width=\columnwidth]{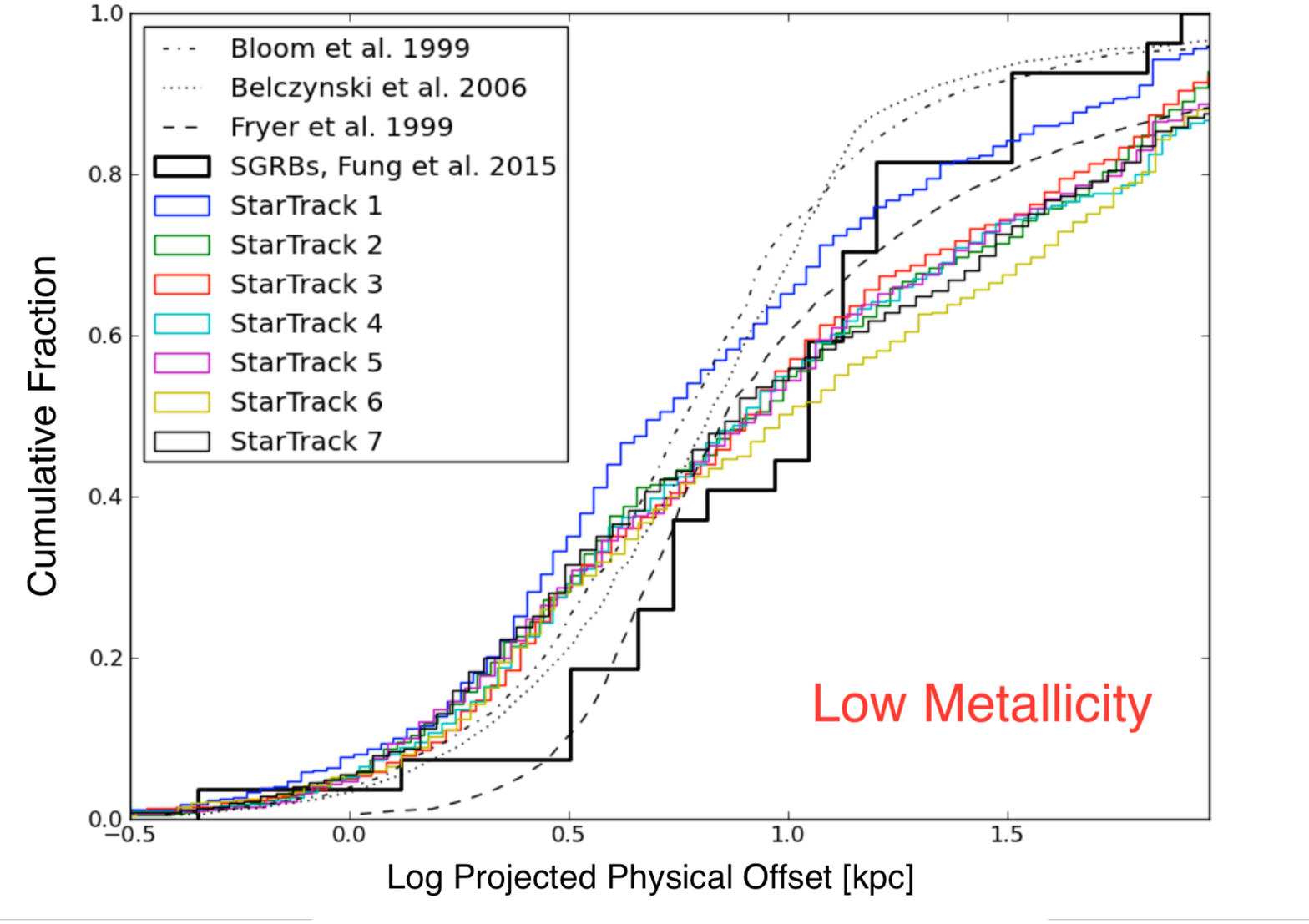}
\caption{Predicted cumulative distributions of projected physical offsets (in kpc) from host galaxies of simulated NS binaries in the cosmological potential. \textit{Top panel:} Offset distributions for 4 velocity kick and delay distributions compared to observations of offsets of sGRBs (blue) and previous simulations (black) in the literature for the older Fryer population synthesis code. \textit{Bottom Panels:} Offset distributions for various kick/delay distributions from the StarTrack population synthesis code (Belczynski et al. 2002, 2008) for low metallicity models (middle panel) and higher metallicity runs (bottom panel).  }
\label{fig:offset}
\end{figure}

Figure~\ref{fig:offset} shows the cumulative offset distribution for the Fryer population synthesis models as well as the low and high-metallicity StarTrack results.  The observed distribution as well as past estimates based on Galactic potential calculations are shown for comparison.  Our full cosmology models predict slightly different distributions than the simplified galactic-potential models of past work (including other galactic potential models such as \cite{voss03}.  We do not produce the larger separations predicted by \cite{zemp09,kelley10}.  But the bigger difference lies in the variance between different population synthesis models and, as the data improves, it is possible that observed offsets can be used to constrain the population synthesis parameters.  All our population synthesis calculations use the same cosmology simulation with one instantiation of the initial density perturbations.  A full uncertainty study would require multiple models of the galaxy evolution.

\subsection{Merger Environments}
\label{sec:environments}

\begin{figure}
\includegraphics[width=\columnwidth]{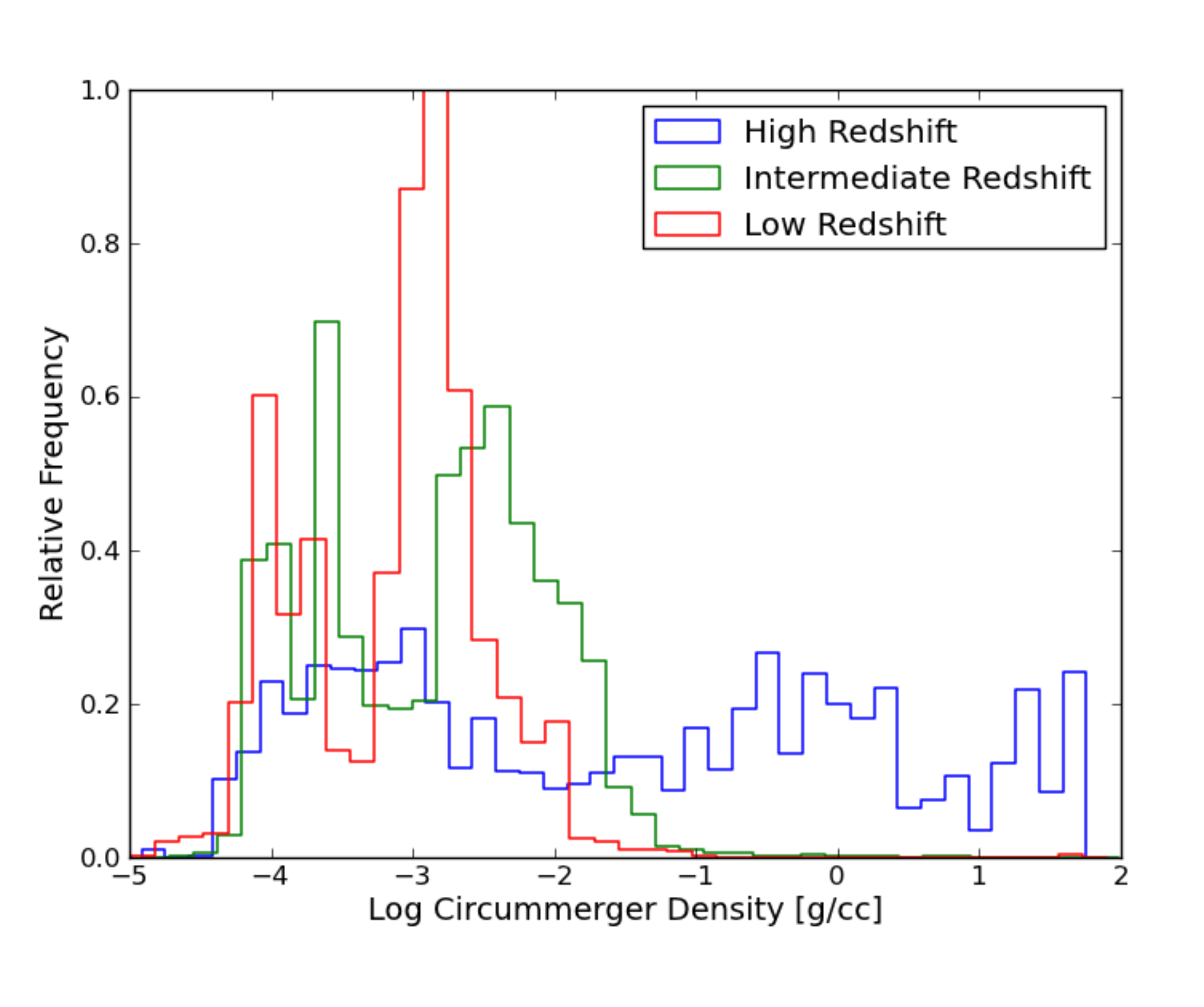}
\includegraphics[width=\columnwidth]{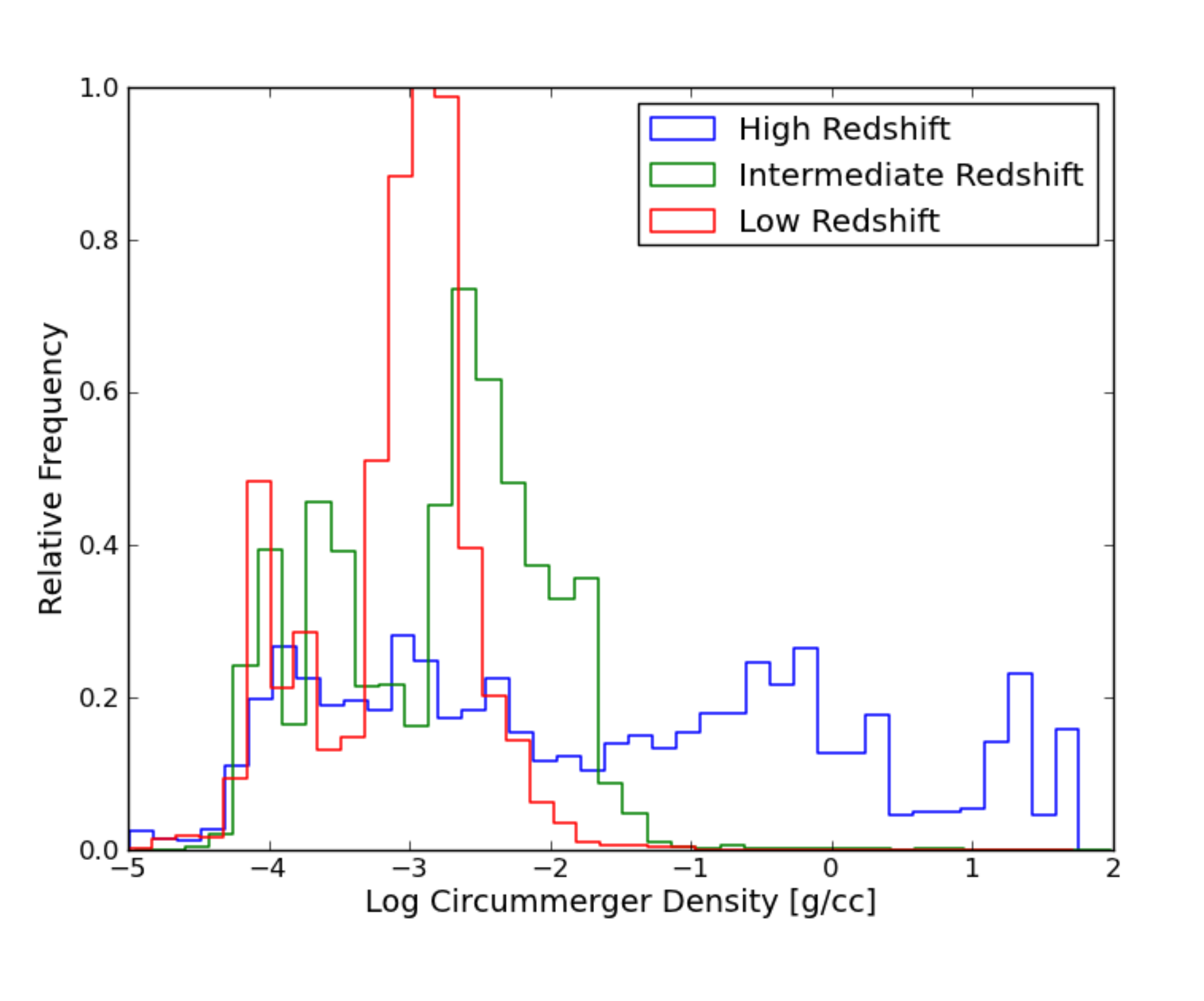}
\includegraphics[width=\columnwidth]{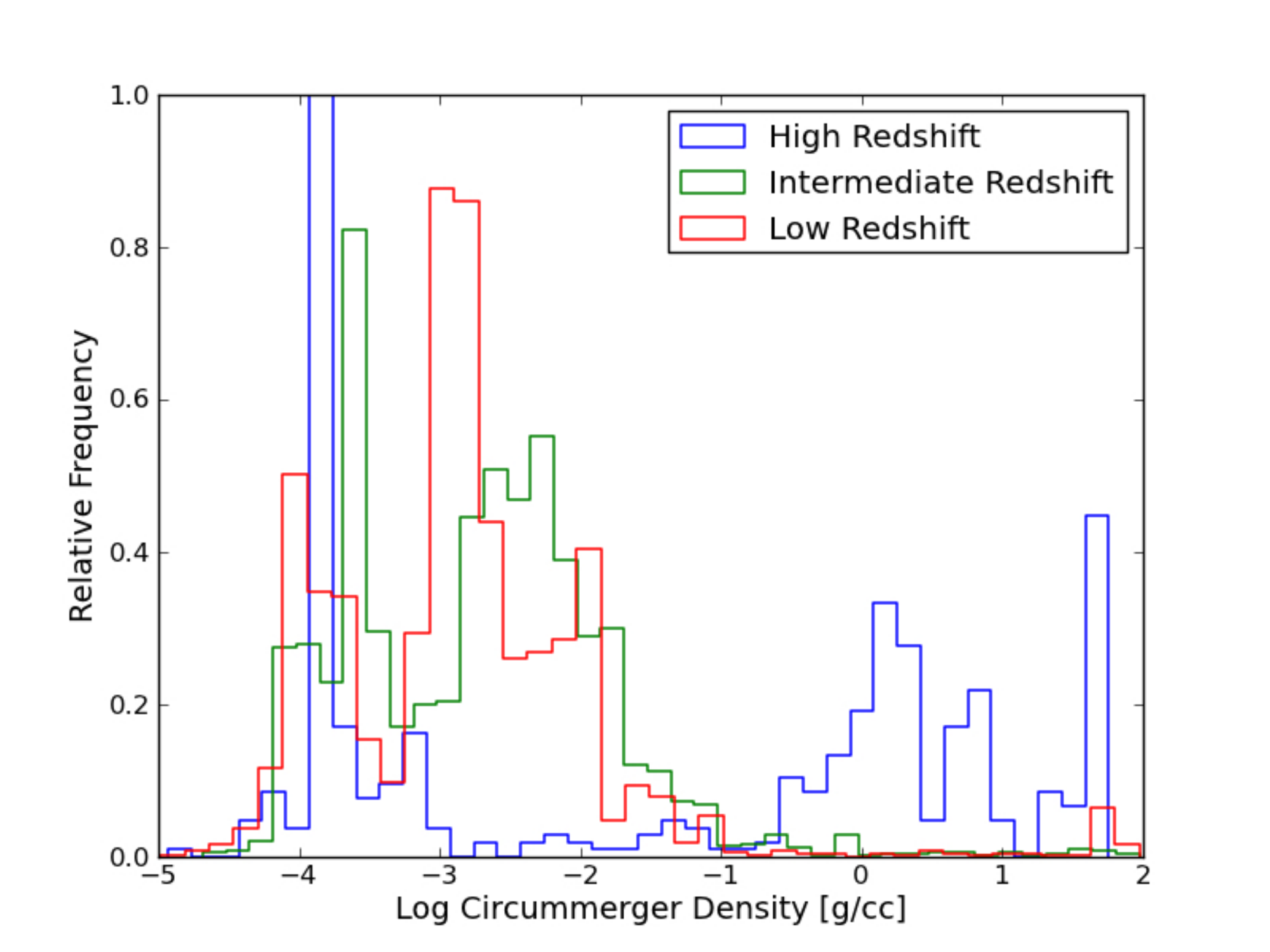}
\caption{Normalized probability distribution functions of DNS circum-merger densities as a function of redshift averaged over all StarTrack high metallicity models in various redshift epochs (High redshift $z > 1$, intermediate redshift: $z \in (0.4, 1.0)$, low redshift: $z < 0.4$). {\it Top Panel:} StarTrack high-metallicity models. {\it Middle Panel:} StarTrack low-metallicity models. {\it Bottom Panel:} population synthesis models from \cite{fryer99}. DNSs are expected to merge at lower densities in the modern universe as star formation slows and binaries wander from their stellar nurseries.}
\label{fig:ST1}
\end{figure}

Our cosmological calculations were able to not only follow the offsets of the merger event to its nearest "host-galaxy" counterpart, but we could also determine the density conditions of these mergers.  Our study uses population synthesis models from~\cite{dominik12} that includes both high- and low-metallicity models.  Because we do not include metallicity evolution, we use these as distinct instantiations for our cosmology runs.  In practice, the low-metallicity models are more valid at high redshift and the high metallicity models are more valid at low redshifts.  Figure~\ref{fig:ST1} shows the distribution of densities in the environment surrounding the merger for the high-metallicity and low-metallicity StarTrack et al. and the Fryer models at 3 different redshift bins: low ($z<0.4$), intermediate ($0.4<z<1.0$) and high ($z>1.0$) redshifts. The densities from our models tend to be, on average, lower than the Milky Way density distribution outlined by \cite{montes16} because many of our models lie outside of their host galaxies.

In all models, the basic trends are that the fraction of mergers occurring in high-density environments is much higher at high-redshifts than at low redshifts.  This argues that the afterglow emission of short duration bursts should be stronger at high redshifts.  This effect is most extreme in the Fryer models that produce very few mergers with short delay times.  As discussed in Section~\ref{sec:popsynth}, these models did not produce mergers in systems undergoing a NS common envelope phase.  The lack of short period binaries means that the binaries in these models travel further from their high-density birth sites before merger.  Hence, there are few mergers from these systems with high densities.  In addition, we see that high-metallicity systems from StarTrack also produced fewer systems with high-density environments than than the low-density models.  This is also caused by the fact that high-metallicity stars have larger radii, producing, on average, binary neutron star systems with larger orbital separations and longer merger times.

\section{Observing Mergers in the Radio}
\label{sec:radio}

GRB observations are able to probe the circum-merger density of the merging system (e.g. Piran et al. 2014; Fong et al. 2015).  The circumburst density profile plays a role in the onset of the afterglow -   in the standard picture of a relativistic external blast wave, the onset of the afterglow occurs around the deceleration time - i.e. when the blast wave has swept up enough external material to begin to decelerate $t_{dec} \propto (E/n)^{1/3}\Gamma^{-8/3}$ \citep{BM76}.  It also affects the strength of the emission and the degree to which the photons are potentially self-absorbed.  
  
The density profile around a long or short GRB also provides a clue to its progenitor.  Long GRBs, believed to arise from the collapse of a massive star, are expected to have wind-like ($1/r^{2}$) density profiles from the mass loss of the star toward the end of its life \citep{CL00}.  If the progenitor of a short GRB is indeed a compact binary merger, on the other hand, these bursts are expected to occur on average in low, constant density environments as they are old stellar populations and migrate out of star forming regions to the outskirts of the galaxy \citep{Pro06,Berger09,Fong10}.  Therefore, it is useful to compare the density inferred or fitted from observations to that predicted from theory of the environments of compact object mergers.  
   

\begin{figure}
\includegraphics[width=\columnwidth]{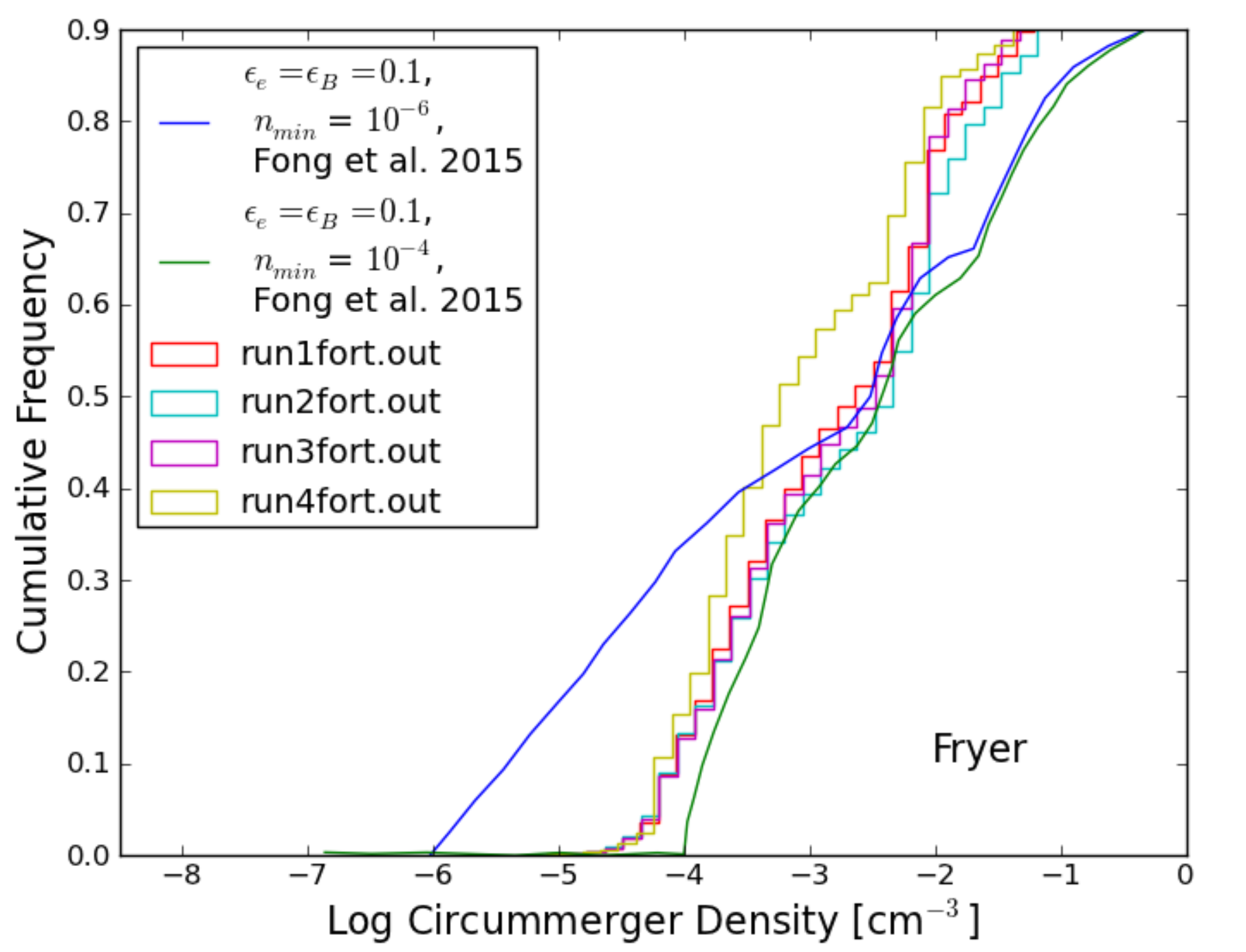}
\includegraphics[width=\columnwidth]{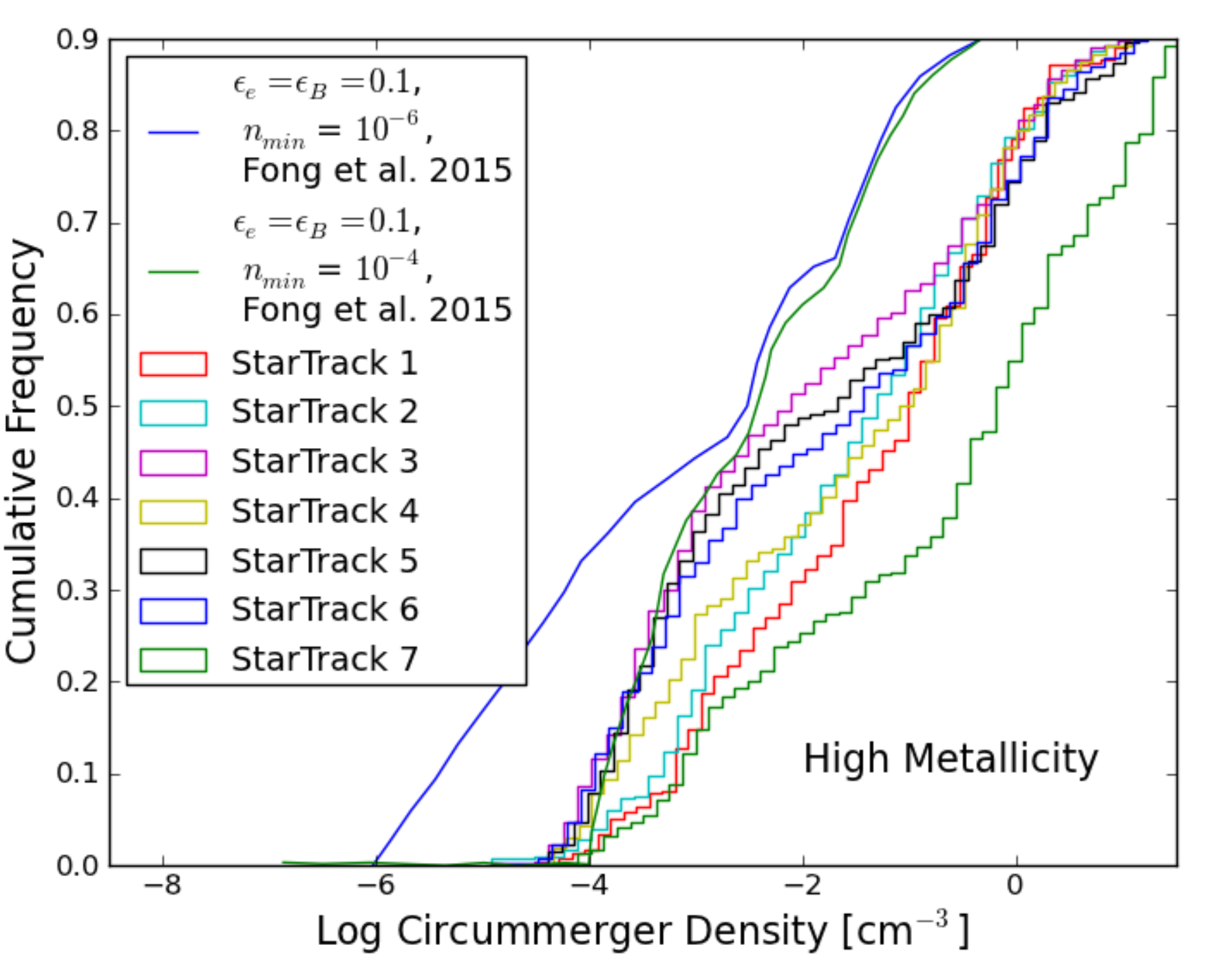}
\includegraphics[width=\columnwidth]{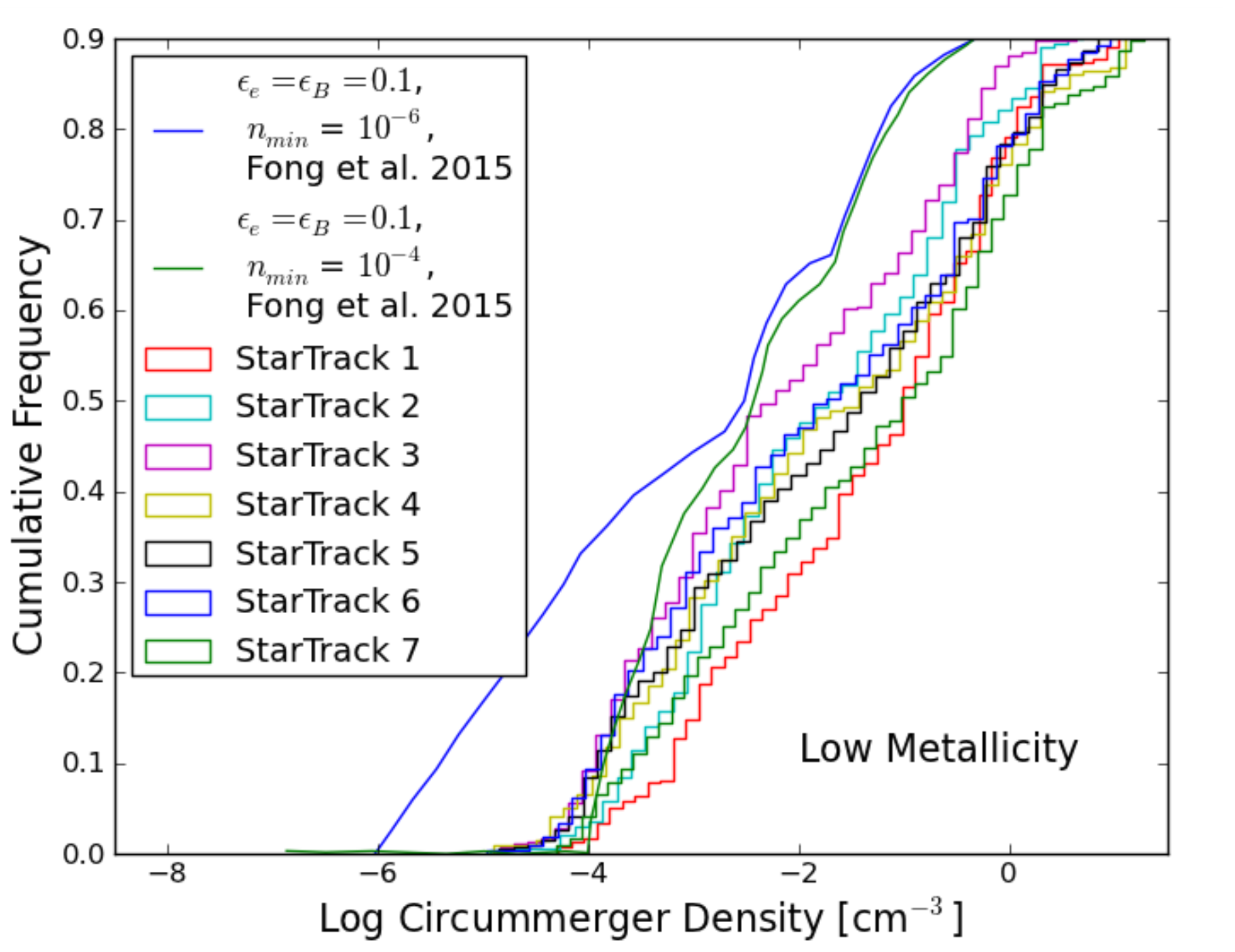}
\caption{Simulated cumulative distribution function of circum-merger density (blue curve) compared with density distributions estimate from Fong et al. 2015 for differing number densities.}
\label{fig:ffong}
\end{figure}

  Figure~\ref{fig:ffong} shows the cumulative circumburst densities obtained from the \citet{fong2015} fits to the GRB afterglow data and those obtained from our cosmological simulations.  Because sGRB observations give rise to circumburst density distributions which are essentially unconstrained on the lower end, the authors choose a minimum density $n_{\mbox{min}} = 10^{-6}$ cm$^{-3}$, a density characteristic of the intergalactic medium, to prevent their sGRB energy and circumburst density distributions from becoming broad. Because the choice of $n_{\mbox{min}}$ is arbitrary, the authors create also create a distribution for $n_{\mbox{min}} = 10^{-4}$ cm$^{-3}$. Both curves describe are integrated from probability distribution functions assuming the microphysics parameters  $\epsilon_e = \epsilon_B = 0.1$ with $\epsilon_e$ and $\epsilon_B$ being the fraction of the energy in electrons the magnetic field respectively.
  Both the fitted and simulated range of densities are similar.  But there is a general trend that StarTrack models produce densities that are higher than the observed densities, with high-metallicity progenitors producing still higher circum-merger densities than their lower metallicity counterparts.  This agrees with results that gamma-ray bursts are typically produced in lower-metallicity systems.  We discuss implications of these density distributions at greater length in \S 5.

As the blast wave plows through the circumburst gas, radio emission originates from the forward shock, the reverse shock and from the interaction the tidally or wind-driven ejected mass with the surrounding medium. Though radio emission is suppressed if the viewing angle is not aligned with the jet, radio flux is also expected for larger inclination angles at later times. The emission flux for each radio component of the sGRB is a strong function of the microphysics parameters $\epsilon_e$, $\epsilon_B$, the electron energy power law index $p$ and the circumburst number density. Many estimates of radio fluxes are arithmetic products of simple functions of these parameters which can make disentangling the effects of a single parameter problematic. Additionally, some of these parameters might not be well-constrained: $\epsilon_e$ and $\epsilon_B$ can in principle vary over orders of magnitude and so fiducial values are frequently chosen. Our study comments on circumburst densities from a first-principles approach as part of an effort to reduce uncertainty in characterizing sGRBs. Flux measurements in all bands combined with the time of the peak of the flux for the different components (forward shock, reverse shock, off-axis emission and tidal ejecta) would help break the parameter degeneracies and help put the density measurements on firmer footing, potentially putting constraints on the formation and evolution of DNS mergers.



\begin{figure}
\includegraphics[width=\columnwidth]{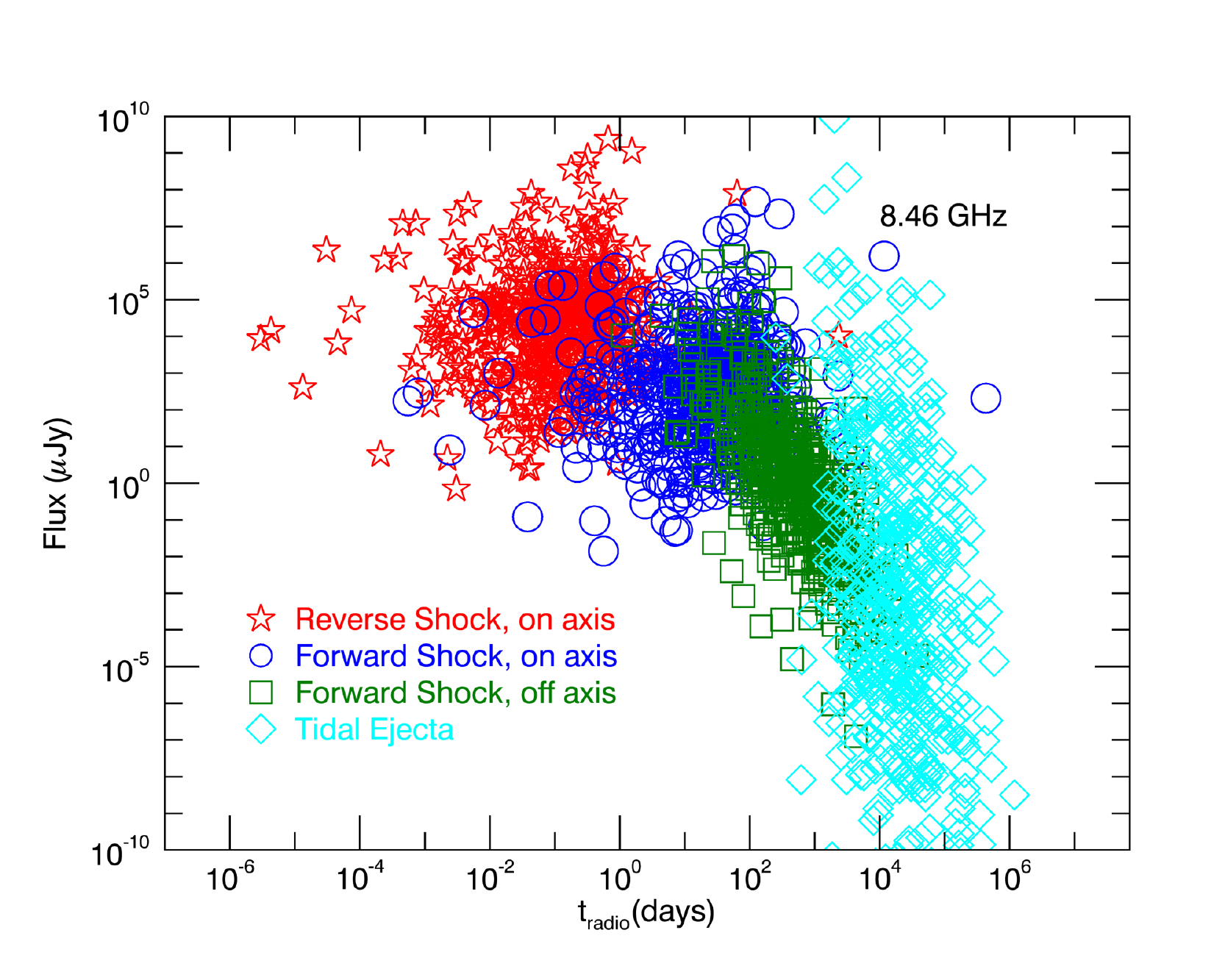}
\includegraphics[width=\columnwidth]{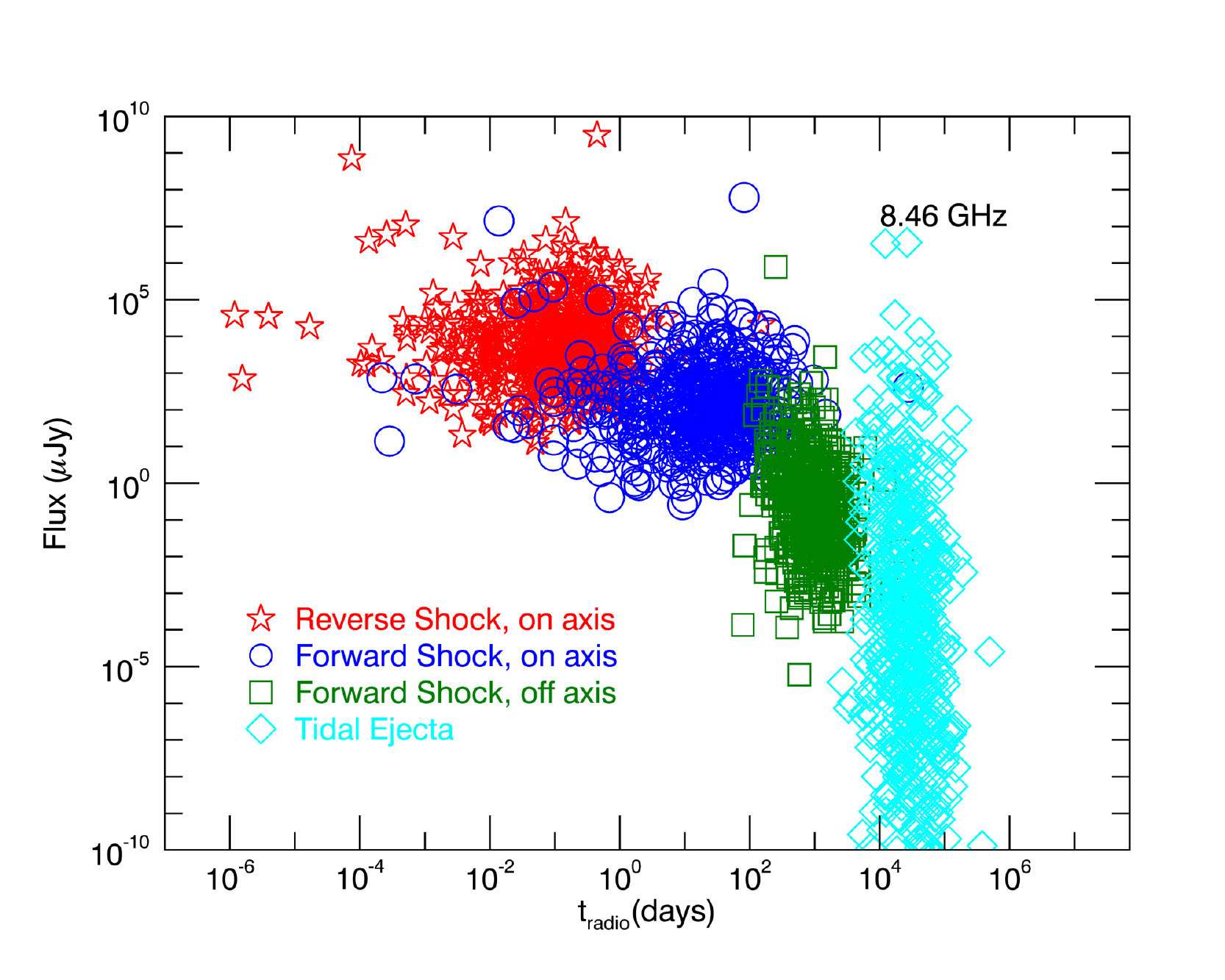}
\caption{Flux as a function of time using density distributions from the run1fort.out calculation (using the \cite{fryer99} models; see Figure~\ref{fig:ST1}) at high redshift (top panel) and low redshifts (bottom panel).  We used a mean energy of $E = 10^{51}$ erg, a mean redshift of $0.5$, a mean value of $p=2.43$. The parameters  $\epsilon_{e}$ and $\epsilon_{B}$ are both drawn from a Gaussian with a mean of 0.1.  The Lorentz factor employed is $\Gamma = 50$.}
\label{fig:Radsig}
\end{figure}

\begin{figure}
\includegraphics[width=\columnwidth]{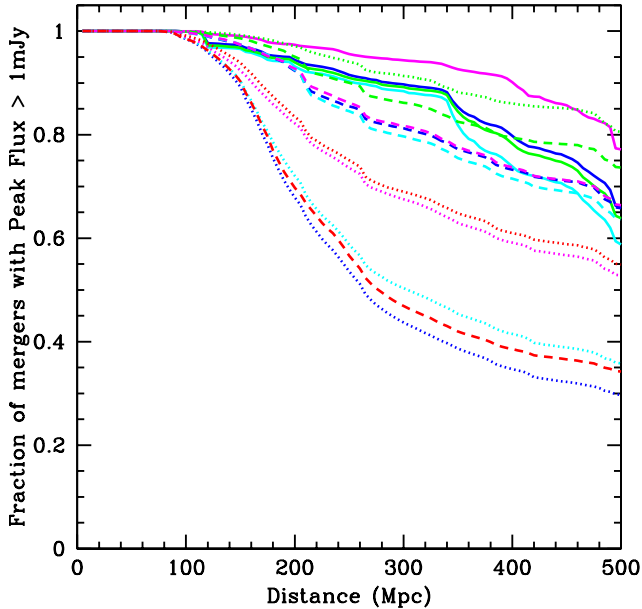}
\caption{Fraction of mergers with radio fluxes at 8.46\,GHz above 1\,mJy as a function of distance.  Here we include 4 models from~\cite{fryer99} (dashed) as well as 5 solar metallicity (dotted) and 5 1/10th solar metallicity (dotted) runs from~\cite{dominik12}.  The fraction of mergers with these high peak fluxes ranges from 35-80\% at 500\,Mpc.}
\label{fig:gwdet}
\end{figure}

Recently, \citet{LR17} examined the different radio emission components of sGRBs, using physical GRB parameters based on the fits to the afterglow spectra by \citet{fong2015} and prescriptions for specific flux for dynamically ejected mass given by \cite{NP11}.
  In Figure ~\ref{fig:Radsig},  we plot the predicted radio fluxes from each of the components (reverse shock, forward shock, off-axis, and tidal ejecta), where densities are taken from the density distribution for the Fryer model run1fort.out which is similar in appearance to the distribution in the top panel of Figure~\ref{fig:ST1}. Data points described as ``on axis'' represent fluxes as measured by an observer looking down the relativistic jet whilst ``off axis'' data points are for fluxes from viewing angles inclined 90$^\circ$ from the jet. The additional sGRB parameters parameters (e.g. energy, fraction of energy in the magnetic and electric field, electron energy index) were taken from distributions that mimic the observed or fitted distributions from \citet{fong2015}. The top panel shows fluxes for circumburst densities at high redshifts and the bottom shows the fluxes for densities at low redshifts. This figure assumes all mergers are observed from $z=0.5$, the average merging redshift of the dataset which better illustrates the relative luminosities of the various radio components of the emission for the distribution of densities in our simulations, leaving the topic of detectability for Figure~\ref{fig:gwdet}. There is a large distribution in flux from the various emission components for our choice of spectral parameters. In principle, we may be able to distinguish these components from the time at which they peak in the radio.  This is relatively well delineated, with the reverse shock peaking earliest, then the forward shock, off-axis emission and finally, the dynamical ejecta component.  

We can use the distribution of peak radio signals to determine the detectability of neutron star mergers detected by advanced LIGO.  Figure~\ref{fig:gwdet} shows the fraction of neutron star mergers with peak radio fluxes at 8.46\,GHz above 1mJy as a function of distance.  This figure includes the results for our different models.  All of these models predict that the radio signal exceed 1mJy out to 100\,Mpc.  But beyond 100\,Mpc, the observed fraction decreases.  The solar metallicity StarTrack and Fryer models have sharper decreases, with as little as 30\% of the systems observed out to 500\,Mpc.  Over 70\% of the low-metallicity models have strong signals out to 500\,Mpc.

\section{Discussion}

In this paper, we coupled cosmology and population synthesis models to determine the distributions of both the neutron star mergers and the densities of their surrounding media.  These properties depend upon assumptions in the population synthesis models and observations of these distributions have the potential to constrain the progenitors of neutron star mergers:  for example, constraints on separations or kick distributions.  The Fryer model assumptions produce wider binaries with lower systematic velocities explain the offset distribution below 10\,kpc, but not peak between 10-30\,kpc (Fig.~\ref{fig:offset}).  StarTrack models that fit the offset distribution peak at 10-30\,kpc tend to overestimate the offset distribution below 10\,kpc.  Despite the differences in binary interactions and kicks, many StarTrack models have similar features to the Fryer model results.  The density distribution can, perhaps, provide a more stringent discriminant.  The Fryer models tend to over-estimate the number sGRBs occurring in dense circumburst densities while the StarTrack models under-estimate this number (Fig.~\ref{fig:ffong}).  The low-kick StarTrack model (model 5) does not produce dramatically different values than the other StarTrack models, and the differences in these predictions lie more in the different treatment of the binary interactions between the two codes (i.e., survivability of common envelope).  Although the data is not strong enough to say anything conclusive, it seems like the right solution lies somewhere in between zero and complete survivability.

A number of uncertainties persist in our calculations.  Our study does not consider dynamically assembled NS binaries in globular clusters or galactic centers \citep[e.g.][]{grindlay06, rod16} but only those formed as zero-age-main-sequence binaries from SFRs in our simulated galaxies. \cite{grindlay06} estimate that $\sim 10-30\%$ of all sGRBs arise from NS binaries formed by progenitor NSs scattering into merging orbits with other NSs in globular clusters. These could represent a lower-kick ($\sigma \lesssim 50$ km/s) population of NS binaries which remain bound to the globular cluster. Our simulations do not have enough resolution to predict circumburst densities for this population NS binaries.

\begin{figure}
\includegraphics[width=\columnwidth]{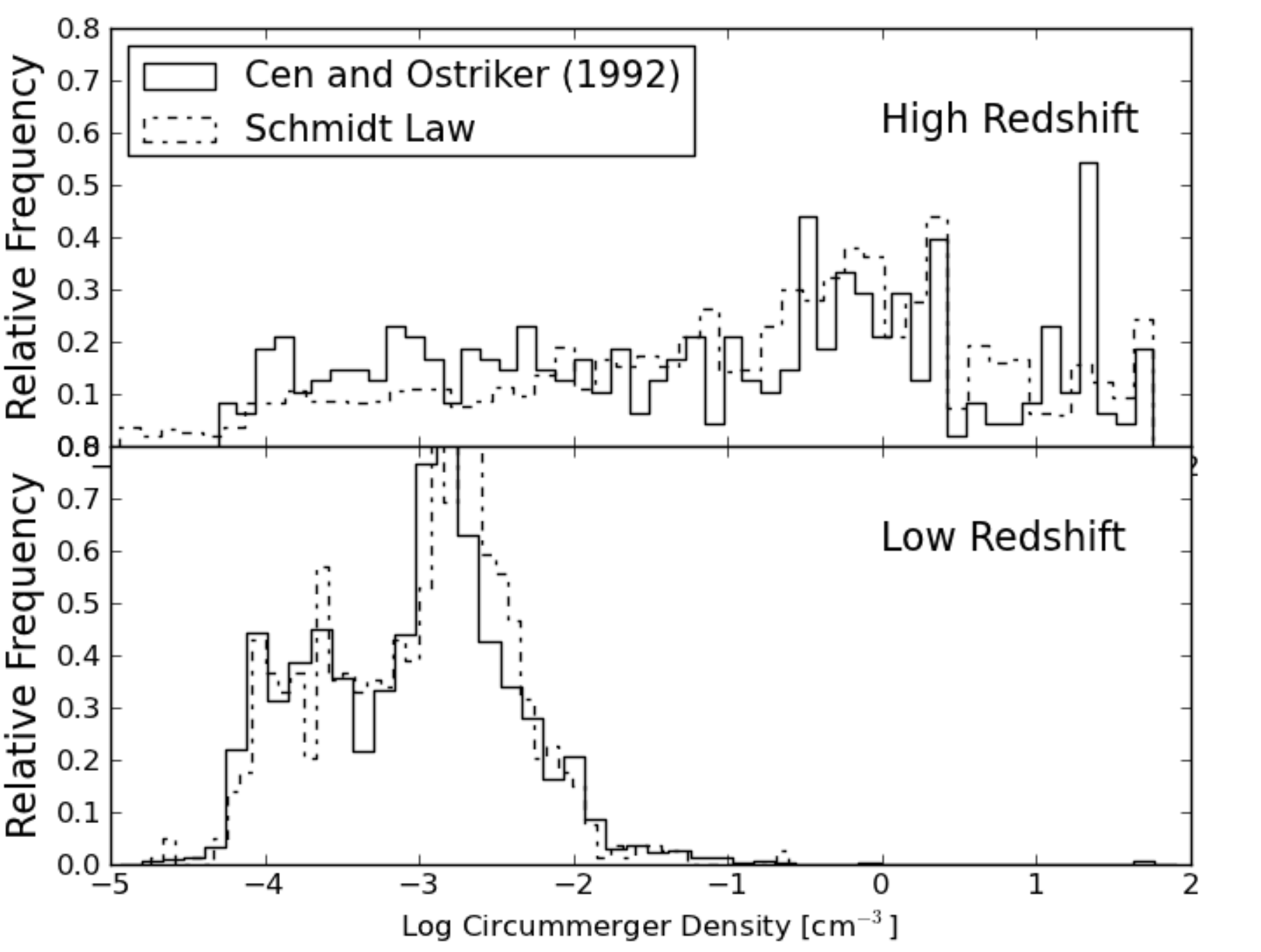}
\caption{Circumburst distributions for the StarTrack 7 run with two different star formation schemes: \cite{cen92} which was used for the calculations in this paper, and the simpler Schmidt Law. \textit{Upper panel:} Predicted circum-merger distributions for mergers occurring before $z = 1$. \textit{Lower panel:} Predicted circum-merger distributions for mergers occurring after $z = 0.4$. }
\label{fig:starformation}
\end{figure}

In this paper, we use the method from \cite{cen92} to calculate SFR. This method produces stars in zones with negative fluid velocity divergence, where the zone's mass exceeds the Jeans mass and where the cooling time is shorter than the dynamical time. Such zones are understood to contain gas which is unstable to collapse and the production of stars in the zone is calculated as a function of the dynamical time.  We reran a post-process adopting a much simpler Schmidt Law \citep{K2003} which estimates the SFR as proportional to the square-root of the local gas density. This scheme predicts binaries arising from new areas of the simulation domain which are not necessarily unstable and so might not be expected to contain SFRs.  Fortunately, our results appear mostly robust between these prescriptions when we reran a post-process with this more primitive scheme (see Figure \ref{fig:starformation}) because natal kicks and the resulting systemic velocities of the NS binaries dominate over specific location in galaxies where the NS binaries form. Particulars regarding the stellar initial mass function (IMF) and how local gas metallicity affect subsequent NS binary formation rates is handled within a given population synthesis model.

This paper demonstrates the potential for merger offsets and density distributions to help us understand binary evolution, but the results depend sensitively on the details and much more data and detailed studies are required to truly constrain this physics.

\acknowledgements
{\bf Acknowledgements:} This project was funded under the auspices of the U.S. Dept. of Energy, and supported by its contract W-7405-ENG-36 to Los Alamos National Laboratory. Simulations at LANL were performed on HPC resources provided under the Institutional Computing program. The work of K.B. was supported from the Polish National Science Center (NCN) grant: Sonata Bis 2 DEC-2012/07/E/ST9/01360. We also thank our anonymous referee whose comments improved this paper.

\bibliographystyle{yahapj}

\begin{thebibliography}{61}
\bibitem[\protect\citeauthoryear{Abbott et al.}{2016}]{Ab16}
Abbott, B.P., et al. 2016, Phys. Rev. Lett, 116, 6
\bibitem[\protect\citeauthoryear{Abbott et al.}{2017}]{Ab17}
Abbott, B.P., et al. 2017, ApJ Lett., 848, L12
\bibitem[\protect\citeauthoryear{Arzoumanian et al.}{2002}]{arzoumanian02}
Arzoumanian, Z., Chernoff, D.F. \& Cordes, J.M. 2002, ApJ, 568, 289
\bibitem[\protect\citeauthoryear{Behroozi et al.}{2014}]{behroozi14}
Behroozi, P.S., Ramirez-Ruiz, E., Fryer, C.L. 2014, ApJ, 792, 123
\bibitem[\protect\citeauthoryear{Belczynski et al.}{2002}]{belczynski02}
Belczynski, K., Bulik, T., Kalogera, V. 2002, ApJ, 571, L147
\bibitem[\protect\citeauthoryear{Belczynski et al.}{2002b}]{belczynski02b}
Belczynski, K., Kalogera, V., Bulik, T. 2002b, ApJ, 572, 407
\bibitem[\protect\citeauthoryear{Belczynski et al.}{2006}]{belczynski06}
Belczynski, K., Perna, R., Bulik, T., Kalogera, V., Ivanova, N., Lamb, D.Q. 2006, ApJ, 648, 1110
\bibitem[\protect\citeauthoryear{Belczynski et al.}{2008}]{belczynski08}
Belczynski, K., Kalogera, V, Rasio, F.A., Taam, R.E., Zezas, A., Bulik, T., Maccarone, T, Ivanova, N. 2008, ApJS, 174, 223
\bibitem[\protect\citeauthoryear{Belczynski et al.}{2017a}]{belczynski17}
Belczynski, K., Ryu, T., Perna, R., Berti, E., Tanaka, T. L., Bulik, T. 2017a, MNRAS, 471, 4702
\bibitem[\protect\citeauthoryear{Belczynski et al.}{2017b}]{belczynski17b}
Belczynski, K., Askar, A., Arca-Sedda, M., Chruslinska, M., Donnari, M., Giersz, M., Benacquista, M., Spurzem, R., Jin, D., Wiktorowicz, G., Belloni, D. astro-ph/1712.00632
\bibitem[\protect\citeauthoryear{Berger et al.}{2005}]{berger05} Berger, E. et al. 2005, Nature, 438, 988
\bibitem[\protect\citeauthoryear{Berger}{2009}]{Berger09}
Berger, E. 2009, ApJ, 690, 231
\bibitem[\protect\citeauthoryear{Berger}{2014}]{Berg14}
Berger, E. 2014, ARA\&A, 52, 43
\bibitem[\protect\citeauthoryear{Bethe}{1998}]{Bethe98}
Bethe, H.A., Brown, G.E. 1998, ApJ, 506, 780
\bibitem[\protect\citeauthoryear{Blandford \& McKee}{1976}]{BM76}
Blandford, R.D. \& McKee, C.F. 1976, Phys.Fluids, 19, 1130
\bibitem[\protect\citeauthoryear{Bloom et al.}{1999}]{bloom99}
Bloom, J.S., Sigurdsson, S., Pols, O.R. 1999, MNRAS, 401, 453
\bibitem[\protect\citeauthoryear{Bodaghee et al.}{2011}]{bod11}
Bodaghee, A., Tomsick, J. A., Rodriguez, J., James, J. B., 2011, 744, 108
\bibitem[\protect\citeauthoryear{Brandt \& Podsiadlowski}{1995}]{brandt95}
Brandt, N., Podsiadlowski, P. 1995, MNRAS, 274, 461
\bibitem[\protect\citeauthoryear{Bryan et al.}{2014}]{bryan14}
Bryan, G. et al. 2014, ApJS, 211, 19
\bibitem[\protect\citeauthoryear{Bulik et al. 1999}{2014}]{bulik99}
Bulik et al. 1999, MNRAS, 309, 629
\bibitem[\protect\citeauthoryear{Cen \& Ostriker}{1992}]{cen92}
Cen, R. \& Ostriker, J. 1992, ApJL, 399, L113
\bibitem[\protect\citeauthoryear{Chang et al.}{2008}]{Chang08}
Chang, P, Spitkovsky, A., Arons, J. 2008, ApJ, 674, 378
\bibitem[\protect\citeauthoryear{Chevalier \& Li}{2000}]{CL00}
Chevalier R.A. \& Li, Z.-Y. 2000, ApJ, 536, 195
\bibitem[\protect\citeauthoryear{Cordes \& Chernoff}{1998}]{cordes98}
Cordes, J.M. \& Chernoff, D.F. 1998, ApJ, 505, 315
\bibitem[\protect\citeauthoryear{D'Avanzo et al. 2009}{2009}]{davanzo09}
D'Avanzo, P. et al. 2009, A\&A, 498, 711
\bibitem[\protect\citeauthoryear{de Mink, S. E. \& Belczynski}{2015}]{DMB2015}
de Mink, S. E., Belczynski, K. 2015, ApJ, 814, 58
\bibitem[\protect\citeauthoryear{Dominik et al.}{2012}]{dominik12}
Dominik, M., Belczynski, K., Fryer, C.L., Holz, D.E., Berti, E., Bulik, T., Mandel, I., O'Shaugnessy, R., 2012, ApJ, 759, 52
\bibitem[\protect\citeauthoryear{Eisenstein \& Hu}{1998}]{eisenstein98}
Eisenstein, D.J. , Hu, W. 1998, ApJ, 496, 605
\bibitem[\protect\citeauthoryear{Fong et al.}{2010}]{Fong10}
Fong, W., Berger, E., \& Fox, D.B., 2010, ApJ, 708, 9
\bibitem[\protect\citeauthoryear{Fong \& Berger}{2013}]{fong13}
Fong, W., \& Berger, E. 776, 18
\bibitem[\protect\citeauthoryear{Fong et al.}{2015}]{fong2015}
Fong, W., Berger, E., Margutti, R., \& Zauderer B.A., 2015, ApJ, 815, 102
\bibitem[\protect\citeauthoryear{Fox et al.}{2005}]{fox05}
Fox, D.B. et al. 2005, Nature, 437, 845
\bibitem[\protect\citeauthoryear{Frail, Waxman, Kulkarni}{2000}]{FWK00}
Frail, D.A., Waxman E., Kulkarni, S.R., 2000, ApJ, 537, 191
\bibitem[\protect\citeauthoryear{Frederiksen et al.}{2004}]{TF04}
Frederiksen, J. Trier et al. 2004, ApJ, 608, L13
\bibitem[\protect\citeauthoryear{Fryer \& Kalogera}{1997}]{fryer97}
Fryer, C.L., Kalogera, V. 1997, ApJ, 489, 244
\bibitem[\protect\citeauthoryear{Fryer et al.}{1998}]{fryer98}
Fryer, C.L., Burrows, A., Benz, W. 1998, ApJ, 496, 333
\bibitem[\protect\citeauthoryear{Fryer et al.}{1999}]{fryer99}
Fryer, C.L., Woosley, S.E., Hartmann, D.H. 1999, ApJ, 526, 152
\bibitem[\protect\citeauthoryear{Fryer \& Kusenko}{2006}]{fryer06}
Fryer, C.L., \& Kusenko, A. 2006, ApJS, 163, 335
\bibitem[\protect\citeauthoryear{Fryer et al.}{2012}]{fryer2012}
Fryer, C.L., Belczynski, K., Wiktorowicz, G. Dominik, M., Kalogera, V., Holz, D. E., ApJ, 749, 91
\bibitem[\protect\citeauthoryear{Gehrels, N. et al.}{2005}]{gehrels05} Gehrels, N. et al. 2005, Nature, 437, 851
\bibitem[\protect\citeauthoryear{Granot \& Sari}{2002}]{GS02}
Granot, J. \& Sari, R. 2002, ApJ, 568, 820
\bibitem[\protect\citeauthoryear{Granot et al.}{2002}]{Gran02}
Granot, J., Panaitescu, A., Kumar, P. \& Woosley, S. 2002, 
\bibitem[\protect\citeauthoryear{Granot \& van der Horst}{2014}]{GvdH14}
Granot, J. \& van der Horst, A.J. 2014, PASA, 31, 8
\bibitem[\protect\citeauthoryear{Grindlay et al.}{2006}]{grindlay06}
Grindlay, J, Zwart, S. P., McMillan, S. 2006, Nature Physics, 2, 116
\bibitem[\protect\citeauthoryear{Guetta \& Piran}{2006}]{GP06}
Guetta, D. \& Piran, T. 2006, A\&A, 453, 823
\bibitem[\protect\citeauthoryear{Hahn \& Abel}{2011}]{hahn11}
Hahn, O., \& Abel, T. MNRAS, 415, 210
\bibitem[\protect\citeauthoryear{Hobbs et al.}{2005}]{hobbs05}
Hobbs, G., Lorimer, D. R., Lyne, A. G., \& Kramer, M. 2005, MNRAS, 360,
974
\bibitem[\protect\citeauthoryear{Hurley et al.}{2000}]{hurley00}
Hurley, J.R., Pols, O.R., Tout, C.A. 2000, MNRAS, 315, 543
\bibitem[\protect\citeauthoryear{Kalogera \& Webbink}{1998}]{kalogera98}
Kalogera, V., \& Webbink, R. F. 1998, ApJ, 493, 351
\bibitem[\protect\citeauthoryear{Kalogera et al.}{2001}]{kalogera01}
Kalogera, V., Narayan, R., Spergel, D. N., Taylor, J. H. 2001, ApJ, 556, 340
\bibitem[\protect\citeauthoryear{Kelley et al.}{2010}]{kelley10}
Kelley, L.Z., Ramirez-Ruiz, E., Zemp, M., Diemand, J., Mandel, I. 2010, ApJL, 725, 91
\bibitem[\protect\citeauthoryear{Kobayashi}{2000}]{Kob00}
Kobayashi, S.2000, ApJ, 545, 807 
\bibitem[\protect\citeauthoryear{Kobayashi \& Zhang}{2003}]{KZ03}
Kobayashi, S. \& Zhang, B. 2003, ApJ, 597, 455 
\bibitem[\protect\citeauthoryear{Kocpac et al.}{2015}]{Kop15}
Kopac, D. 2015, ApJ, 806, 179
\bibitem[\protect\citeauthoryear{Kouveliotou et al.}{1993}]{kouveliotou93}
Kouveliotou, C., Meegan, C.A., Fishman, G.J., Bhat, N.P., briggs, M.S., Koshut, T.M, Paciesas, W.S., Pendleton, G.N. 1993, ApJ, 413, L101 
\bibitem[\protect\citeauthoryear{Kim et al.}{2003}]{kim03}
Kim, C., Kalogera, V., Lorimer, D. R., ApJ, 584, 985
\bibitem[\protect\citeauthoryear{Kulkarni, Frail \& Sari}{1999}]{KFS99}
\bibitem[\protect\citeauthoryear{Kravtsov}{2003}]{K2003}
Kravtsov, A. V. 2003, ApJ, 590, L1-L4
\bibitem[\protect\citeauthoryear{Laskar et al.}{2013}]{Las13}
Laskar, T. et al. 2013, ApJ, 776, 119
\bibitem[\protect\citeauthoryear{Laskar et al.}{2016}]{Las16}
Laskar, T. et al. 2016, ApJ, submitted, arXiv1606.08873
\bibitem[\protect\citeauthoryear{Lee \& Ramirez-Ruiz}{2007}]{Lee07}
Lee, W.H., Ramirez-Ruiz, E. 2007, New Journal of Physics, 9, 17
\bibitem[\protect\citeauthoryear{Lemoine}{2013}]{Lem13}
Lemoine, M. 2013, MNRAS, 428, 845
\bibitem[\protect\citeauthoryear{Li \& Paczynski}{1998}]{li98}
Li, L.-X., Paczynski, B. 1998, ApJ, 507, L59
\bibitem[\protect\citeauthoryear{Lloyd \& Petrosian}{2000}]{LP00}
Lloyd, N.M. \& Petrosian, V. 2000, ApJ,  543, 722
\bibitem[\protect\citeauthoryear{Lloyd-Ronning et al.}{2017}]{LR17}
Lloyd-Ronning, N.M., Fryer, C.L., Hartmann, D.H., Wiggins, B. 2017, chapter in {\em ngVLA Science Book}, arXiv 1709.08512
\bibitem[\protect\citeauthoryear{Macleod \& Ramirez-Ruiz}{2015}]{macleod15}
Macleod, M., \& Ramirez-Ruiz, E. 2015, ApJ, 798, 19
\bibitem[\protect\citeauthoryear{Meszaros \& Rees}{1997}]{MR97}
Meszaros, P. \& Rees, M.J. 1997, ApJ, 476, 231
\bibitem[\protect\citeauthoryear{Milosavljevic \& Nakar}{2006}]{MN06}
Milosavljevic, M. \& Nakar, E. 2006 ApJ, 641, 978
\bibitem[\protect\citeauthoryear{Montes et al.}{2016}]{montes16}
Montes, G., Ramirez-Ruiz, E., Naiman, J., Shen, S., Lee, W.H. 2016, ApJ, 830, 12
\bibitem[\protect\citeauthoryear{Nakar \& Piran}{2005}]{NP05}
Nakar, E. \& Piran, T. 2005, ApJ, 619, L147
\bibitem[\protect\citeauthoryear{Nakar \& Piran}{2011}]{NP11}
Nakar, E. \& Piran, T. 2011, Nature, 478, 82
\bibitem[\protect\citeauthoryear{Nakar, Piran, \& Granot}{2002}]{NPG02}
Nakar, E., Piran, T. \& Granot, J. 2002, ApJ, 579, 699
\bibitem[\protect\citeauthoryear{Nakar, Piran, \& Rosswog}{2013}]{NPR13}
Nakar, E., Piran, T. \& Rosswog, S. 2013, MNRAS, 430, 2121
\bibitem[\protect\citeauthoryear{Oren et al.}{2004}]{Oren04}
Oren, 2004, MNRAS, 353, L35
\bibitem[\protect\citeauthoryear{O'Shaughnessy et al.}{2008}]{Osh08}
O'Shaughnessy, R., Kim, C., Kalogera, V., Belczynski, K. 2008, ApJ, 672, 479
\bibitem[\protect\citeauthoryear{Pelletier et al.}{2017}]{Pell17}
Pelletier, G., Bykov, A., Ellison, D., Lemoine, M. 2017, Space Science Reviews; arXiv 1705.05549
\bibitem[\protect\citeauthoryear{Pfahl et al.}{2002}]{pfahl02}
Pfahl, E.,Rappaport, S., Podsiadlowsk, P., 2002, ApJ, 573, 283
\bibitem[\protect\citeauthoryear{Phinney}{1991}]{Phi91}
Phinney, E.S. 1991, ApJ, 380, L17
\bibitem[\protect\citeauthoryear{Piran}{2004}]{Pir04}
Piran, T. 2004, RvMP, 76, 1143
\bibitem[\protect\citeauthoryear{Piran et al.}{2014}]{Pir14}
Piran, T., Korobkin, O., Rosswog, S. 2014, arXiv preprint arXiv:1401.2166
\bibitem[\protect\citeauthoryear{Popham et al.}{1999}]{popham99}
Popham, R., Woosley, S.E., Fryer, C. 1999, ApJ, 518, 356
\bibitem[\protect\citeauthoryear{Portegies Zwart \& Yungelson}{1998}]{Por98}
Portegies Zwart, S. F., Yungelson, L.R. 1998, A\&A, 332, 173 
\bibitem[\protect\citeauthoryear{Prochaska}{2006}]{Pro06}
Prochaska, J.X., et al. 2006, ApJ, 642, 989
\bibitem[\protect\citeauthoryear{Resmi \& Zhang}{2016}]{RZ16}
Resmi, L. \& Zhang, B. 2016, ApJ, 825, 48
\bibitem[\protect\citeauthoryear{Rezolla et al.}{2010}]{Rez10}
Rezolla, L., Baiotii, L., Giacomazzo, B., Link, D., Font, J.A. 2010, Classical Quantum Gravity, 27, 114105
\bibitem[\protect\citeauthoryear{Rhoads}{1997}]{Rhds97}
Rhoads, J.E. 1997, ApJ, 478, L1
\bibitem[\protect\citeauthoryear{Rodriguez}{2016}]{rod16}
Rodriguez, C. L., Chatterjee, S., Rasio, F. A. 2016, Phys. Rev. D, 93, id.084029
\bibitem[\protect\citeauthoryear{Rosswog et al.}{1999}]{Ros99}
Rosswog, S., et al. 1999, A\&A, 341, 499
\bibitem[\protect\citeauthoryear{Ruffert \& Janka}{2001}]{RJ01}
Ruffert, M. \& Janka, H.-T.  2001, A\& A 380, 544
\bibitem[\protect\citeauthoryear{Sari \& Piran}{1999}]{SP99}
Sari,R. \& Piran, T. 1999, ApJ, 520, 641
\bibitem[\protect\citeauthoryear{Sironi, Keshet, \& Lemoine}{2015}]{SKL15}
Sironi, L., Keshet, U. \& Lemoine, M. 2015, SSSRv, 191 579
\bibitem[\protect\citeauthoryear{Skory et al.}{2010}]{skory10}
Skory, S., Turk, M.J., Norman, M.L., Coil, A.L 2010, ApJS, 191, 43
\bibitem[\protect\citeauthoryear{Soderberg \& Ramirez-Ruiz}{2003}]{SRR03}
Soderberg, A. \& Ramirez-Ruiz, E. 2003, MNRAS, 345, 854
\bibitem[\protect\citeauthoryear{Soderberg et al.}{2006}]{Sod06}
Soderberg, A. et al. 2006, ApJ, 638, 930
\bibitem[\protect\citeauthoryear{Stockem Novo et al.}{2015}]{SN15}
Stockem Novo, A. et al. 2015, ApJ, 803, L29
\bibitem[\protect\citeauthoryear{Tanvir et al.}{2013}]{tanvir13}
Tanvir, N.R., Levan, A.J, Fruchter, A.S., Hjorth, J., Hounsell, R.A., Wiersema, K., Tunnicliffe, R.L. 2013, Nature, 500, 547
\bibitem[\protect\citeauthoryear{Troja et al.}{2008}]{troja08}
Troja, E., King, A.R., O'Brien, P.T., Lyons, N., Cusumano, G. 2008, MNRAS, 385, L10
\bibitem[\protect\citeauthoryear{vanEerten et al.}{2011}]{vanE11}
van Eerten, H.J., MacFadyen, A.I.,  \& Zhang, W. 2011, AIPC, 1358, 159
\bibitem[\protect\citeauthoryear{Voss \& Tauris}{2003}]{voss03}
Voss, R., Tauris, T.M 2003, MNRAS, 342, 1169
\bibitem[\protect\citeauthoryear{Waxman}{2004}]{Wax04}
Waxman, E. 2004, ApJ, 602, 886
\bibitem[\protect\citeauthoryear{Woosley}{1993}]{woosley93}
Woosley, S. E. 1993, ApJ, 405, 273
\bibitem[\protect\citeauthoryear{Xu \& Li}{2010}]{xu10}
Xu, X.-J., Li, X.-D. 2010, ApJ, 716, 114
\bibitem[\protect\citeauthoryear{Yamamoto, Shibata, \& Taniguchi}{2008}]{YST08}
Yamamoto, T., Shibata, M. \& Taniguchi, K. 2008, Phys. Rev. D. 78, 064054
\bibitem[\protect\citeauthoryear{Zemp et al.}{2009}]{zemp09}
Zemp, M., Ramirez-Ruiz, E., Diemand, J. 2009, ApJ, 705, 186
\bibitem[\protect\citeauthoryear{Zhang \& Yan}{2011}]{ZY11}
Zhang, B., Kobayashi, S. \& Meszaros, P. 2003, ApJ, 595, 950
\bibitem[\protect\citeauthoryear{Zou et al.}{2005}]{ZWD05}
Zou, Y.C, Wu, X.F. \& Dai, Z.G. 2005, MNRAS, 363, 93
\end{thebibliography}

\end{document}